\documentclass[12pt]{amsart} 
\usepackage[margin=2.5cm]{geometry}

\usepackage[square,numbers,sort&compress]{natbib}

\makeatletter
\renewcommand\subsection{\@startsection{subsection}{2}
  \z@{-.5\linespacing\@plus-.7\linespacing}{.5\linespacing}
  {\normalfont\scshape}}
\renewcommand\subsubsection{\@startsection{subsubsection}{3}
  \z@{.5\linespacing\@plus.7\linespacing}{-.5em}
  {\normalfont\scshape}}
\def\paragraph{\@startsection{paragraph}{4}
  \z@\z@{-\fontdimen2\font}
  {\normalfont\itshape}}
\makeatother

\usepackage{setspace} \usepackage{enumerate} \usepackage{amssymb}  
 \usepackage{bm,enumerate,amsthm,url,mathabx}
 \usepackage[normalem]{ulem}
 \usepackage{xcolor}
 \usepackage{graphicx}

\theoremstyle{remark}  
\newtheorem{remark}{Remark}[section]
\newtheorem*{remark*}{\textcolor{red}{Remark}}

\theoremstyle{remark}  
\newtheorem{proposition}{Proposition}[section]
\newtheorem*{proposition*}{Proposition}

\theoremstyle{remark}  
\newtheorem{corollary}{Corollary}[section]
\newtheorem*{corollary*}{Corollary}

\theoremstyle{remark}  
\newtheorem{definition}{Definition}[section]

\theoremstyle{remark}  
\newtheorem{example}{Example}[section]

\title{Absolutely No Free Lunches!}
 \author{Gordon Belot}\thanks{For helpful comments and discussion, thanks to: an anonymous referee, Josh Hunt, Thomas Icard, Mikayla Kelley, Tom Sterkenburg, and Bas van Fraassen, and Francesca Zaffora Blando. \medskip}
 \address{Department of Philosophy, University of Michigan \hfill belot@umich.edu}

\begin{document}

\begin{abstract}
This paper is concerned with learners who aim to learn patterns in infinite binary sequences: shown longer and longer initial segments of a binary sequence, they either attempt to predict whether the next bit will be a 0 or will be a 1 or they issue forecast probabilities for these events. Several variants of this problem are considered. In each case, a no-free-lunch result of the following form is established: the problem of learning is a formidably difficult one, in that no matter what method is pursued, failure is incomparably more common that success; and difficult choices must be faced in choosing a method of learning, since no approach dominates all others in its range of success. In the simplest case, the comparison of the set of situations in which a method fails and the set of situations in which it succeeds is a matter of cardinality (countable vs. uncountable); in other cases, it is a topological matter (meagre vs. co-meagre) or a hybrid computational-topological matter (effectively meagre vs. effectively co-meagre). 
\end{abstract}

\keywords{Induction, Learning, Extrapolation, Forecasting, No-Free-Lunch Theorems\medskip}

\maketitle

\section{Introduction} \label{secIntro} \thispagestyle{empty}

The various \emph{no-free-lunch theorems} of statistical, computational, and formal learning theory offer ways to make precise the basic insight that there can be no optimal general-purpose approach to learning. These theorems come in two main forms.  Some show that there are contexts in which certain approaches to learning succeed in each salient situation, but that each such approach has the same expected performance across those possible situations.\footnote{For results of this kind, see \citep{wolpert1997no,wolpert2002supervised,ho2002simple}. For further discussion, see \citep{von2011statistical} and \citep{Sterkenburg:2020aa}.\label{fnMacCready}} Results of this kind are  \emph{measure-relative}: in order for expectations to be defined, a measure must be imposed on the space of situations that a learner might face---and the results in question only hold relative to some of the measures that one might impose \citep{rao1995every}. Results of a second kind are \emph{absolute} in the sense that they do not rely upon the choice of a measure on the space of envisaged situations. Here are  descriptions of two paradigmatic results of this kind. 
\begin{quote}
Maybe there exists some kind of universal learner, that is, a learner who has no prior knowledge about a certain task and is ready to be challenged by any task? \ldots\ The no-free-lunch theorem states that no such universal learner exists. To be more precise, the theorem states that for binary classification prediction tasks, for every learner there exists a distribution on which it fails. \ldots\ In other words, the theorem states that no learner can succeed on all learnable tasks---every learner has tasks on which it fails while other learners succeed.\footnote{Shalev--Shwartz and Ben--David \citep[p. 36]{shalev2014understanding}.   \\ \smallskip  \hfill Forthcoming in \emph{Theoretical Computer Science.}} \vspace{1.5mm} \\
Let T be any learning machine\ldots.  [W]e will \emph{defeat} the machine T. That is, we will have constructed a regularity, depending on T, which is beyond the power of T to extrapolate. However \ldots\ it is always possible to build another machine which can extrapolate every regularity that T can extrapolate and also extrapolate the one that T \emph{can't} extrapolate. Thus, there cannot exist a cleverest learning machine: for, for every learning machine T, there exists a machine T$^\prime$ which can learn everything that T can learn and more besides.\footnote{Putnam \citep[pp. 6 f.]{putnam1964probability}. As noted by Case and Smith \citep[p. 208]{case1983comparison}: ``This appears to be the earliest result indicating that there may be some difficulty with the mechanization of science.'' See also \citep[Theorem I.5]{gold1967language}.}  
\end{quote}
We can think of such results as encapsulating two facts about the predicament of learners situated in certain contexts. (a) They face a daunting problem: no approach they might adopt succeeds across all envisaged situations. (b) Difficult choices must be made: different approaches succeed in different situations, with no approach dominating all others in its range of success.

Here we assemble some more or less elementary results, some already well-known, that combine to give no-free-lunch results of the second, absolute variety, applicable to agents attempting to learn patterns in binary data streams.\footnote{A problem that, according to Li and Vit\'anyi \citep[p. 6]{li2013introduction}, ``constitutes, perhaps, the central task of inductive reasoning and artificial intelligence.''} Nature presents our agents with one-way infinite binary sequences one bit at a time and after each bit is revealed each agent is asked to make a prediction about the next bit. We will consider five models of learning, differing from one another as to what sort of predictions our agents are required to make or as to the criterion of success. And for each model, we will consider variants in which neither the agent nor Nature is required to follow a computable strategy, in which the agent is required to follow a computable strategy but Nature is not, and in which both the agent and Nature are required to follow computable strategies. For each variant of each model, we establish both elements required for a no-free-lunch result. (i) Difficult choices must be faced in selecting a method of learning: we will show that no approach dominates all of its rivals, either by showing that for each method there is another that succeeds in a disjoint set of situations (\emph{evil twin results}) or by showing that for every method there is another that succeeds in a strictly larger family of situations (\emph{better-but-no-best results}). (ii) We also show our learners face a formidably difficult problem: for each of the problems we consider, there is a sense in which for any method of addressing that problem, the situations in which it fails are incomparably more common than the situations in which it succeeds.\footnote{So here we go beyond the paradigm results mentioned above, which show only that for each method, there exists a situation in which it fails. For other results of the sort developed here, see \citep{fortnow1998relative}.} 

Following some preliminaries in Section \ref{secChar}, we investigate in Section \ref{secLearnSeq} the predicament of learners who must attempt to guess, before each bit is revealed, whether it will be a 0 or a 1.\footnote{For early investigations of such agents, see the papers of Putnam \citep{putnam1963degree,putnam1964probability} and Gold  \citep{gold1967language}.\label{fnGold}} We will consider two criteria of success for such \emph{next-value} learners: when facing a given data stream they should eventually predict each new bit correctly (NV-learning); or the should predict each new bit correctly, except for a family of errors that has vanishing asymptotic density (weak NV-learning).\footnote{The notion of NV-learning is due to B\={a}rzdi\c{n}\v{s} \citep{barzdins1972prognostication}  (see also Blum and Blum \citep{blum1975toward}). The notion of weak NV-learning is due to Podnieks and Kinber \citep[pp. 80 f.]{Podnieks:1974aa}.}  In Section \ref{subsecNV} we will see that for NV-learning of arbitrary sequences, failure is incomparably more common than success in the sense that any method for predicting bits succeeds for a countable family of binary sequences and fails for an uncountable family of binary sequences. In Section \ref{subsecNVw} we will see that for weak NV-learning of arbitrary sequences, any method succeeds for an uncountable set of sequences and fails for an uncountable set of sequences, but the successes are always incomparably less common than the failures in a topological sense, forming a meagre set. In Section \ref{subsecCompSeq}, we restrict attention to computable methods for the next-value learning of computable sequences and find that  for any method, the sets of success and failures are equivalent both from the point of view of cardinality and the point of view of topology---but that the successes are nonetheless incomparably less common than the failures in the hybrid topological-computational sense (due to Mehlhorn \citep{mehlhorn1973size}) that they form an effectively meagre set.\footnote{In the case of NV-learning, this result is due to Fortnow \emph{et al.} \citep{fortnow1998relative}.} Along the way we will  see that the notion of weak NV-learning, while strictly weaker than the notion of NV-learning, is neither weaker nor stronger than two other variants of NV-learning, NV\hspace{1pt}$^\prime$-learning (due to B\={a}rzdi\c{n}\v{s} \citep{barzdins1972prediction}) and NV\hspace{1pt}$\hspace{1pt}^{\prime\prime}$-learning (due to Podnieks \citep{Podnieks:1974aa}). 

In Section \ref{secLearnMeas} we turn to agents who face a data stream sampled from a probability measure chosen by Nature and who are required to issue forecast probabilities for the next bit's being a 0 or a 1 just before it is revealed.\footnote{This model of learning appears in Solomonoff \citep{solomonoff1964formal}.} We consider three criteria of success for agents engaged in such \emph{next chance} prediction: we can ask that for any event, the probabilities that our agents assign to that event converge almost certainly to the true probability as they see larger and larger data sets (strong NC-learning);  we can ask that their forecast probabilities for the next bit become arbitrarily accurate, almost certainly, in the limit of large data sets (NC-learning); or we can ask that they meet the last-mentioned standard \emph{modulo} a set of missteps of vanishing asymptotic density (weak NC-learning).\footnote{These criteria of success were introduced by Blackwell and Dubins \citep{blackwell1962merging}, Kalai and Lehrer \citep{kalai1994weak}, and Lehrer and Smorodinsky \citep{lehrer1996merging}. The criteria of success employed in the literature on Solomonoff induction differ in focussing on average or expected performance in the long run---on the relation between those notions and the notion of NC-learning, see \citep{ryabko2007sequence}.} For the problem of next-chance learning in the face of a data stream generated by an arbitrary measure, we see in Sections \ref{subsecStrongNC} and \ref{subsecWeakMeas} that for any of our criteria of success, each method fails for an uncountable set of measures that Nature might have chosen and succeeds for an uncountable set of such measures---but that the former set is always incomparably smaller than the latter, being meagre. In \ref{secComp}, we restrict attention to computable strategies for next-chance learning in contexts in which the data stream is generated by a computable measure and find, for each of our three criteria of success, that the set of learnable measures is an effectively meagre subset of the family of computable measures.  Section \ref{secDiscuss} provides a few concluding remarks.

\section{Preliminaries} \label{secChar}

\subsection{The Main Characters} \label{subsecMainChar}
We will be concerned below with a number of topological spaces. \begin{enumerate}[(i)]

\item The space of bits, $\mathcal{B}:= \{0,1 \},$ equipped with the discrete topology (so every subset of $\mathcal{B}$ is open). 

\item Finite products of $\mathcal{B}$ with itself: for each $n\in \mathbb{N},$ the space of $n$-bit strings, $\mathcal{B}^n,$ equipped with the discrete topology (we count $0$ as a natural number and use $\varnothing$ to denote either  the empty string of zero bits that is the sole member of $\mathcal{B}^0$ or the empty set, depending on context).  We will think of elements of $\mathcal{B}^n$ as strings (concatenations of symbols) rather than as $n$-tuples. For $w\in \mathcal{B}^n$ and $m\leq n$ we write $w(m)$ for the $m$th bit of $w$ and write $w[m]$ for the $m$-bit initial segment of $w.$ For $m,n\in \mathbb{N}$ with $m\leq n,$ we have the natural projection map $\pi_{nm}: w\in \mathcal{B}^n \mapsto w[m]\in \mathcal{B}^m.$

\item The space of binary strings, $\mathcal{B}^* := \bigcup_{n=0}^\infty \mathcal{B}^n,$ also equipped with the discrete topology. If $v$ and $w$ are binary strings we write $v.w$ for the string that results from concatenating $v$ and $w$ (in that order) and write $v.w^2$ for the results of concatenating $v$ with $w$ and with $w,$ etc. We write $|w|$ for the number of bits in binary string $w.$ 

\item \emph{Cantor space,} $\mathcal{C},$ the set of all infinite binary sequences equipped with the product topology (we take sequences to be indexed by positive natural numbers). We can characterize this topology as follows: if $w$ is an $n$-bit string, then we use $B_w$ to denote the set of sequences whose first $n$ bits are given by $w$; the set of all such $B_w$ (as $w$ ranges over $\mathcal{B}^*$) is a basis for the product topology and we call the $B_w$ \emph{basic open sets.} \emph{Illustration}: the set of sequences that have 0 as their second bit is an open set because it is the union of the basic open sets $B_{00}$ and $B_{10}.$ For $\sigma\in \mathcal{C}$ we write $\sigma(m)$ for the $m$th bit of $\sigma$ and write $\sigma[m]$ for the $m$-bit string formed by concatenating the first $m$ bits of $\sigma.$ For each $n \in \mathbb{N}$ we have the natural projection map $\pi_n: \sigma\in \mathcal{C} \mapsto \sigma[n]\in \mathcal{B}^n.$ A sequence of points  $\sigma_1,$ $\sigma_2,$ $\sigma_3,$ \ldots\ in Cantor space converges to $\sigma \in \mathcal{C}$ if and only if for each $k,$ there exists an $N$ so that for $n\geq N,$ $\sigma_n(k)=\sigma(k).$ We use $\mathfrak{B}$ to denote the $\sigma$-algebra of Borel subsets of $\mathcal{C}.$ We use $\mathfrak{C}$ to denote the subspace of $\mathcal{C}$ consisting of computable sequences (i.e., the $\sigma$ such that the map $k\in \mathbb{N} \mapsto \sigma[k]$ is computable).

\item For each $k\in \mathbb{N},$ the space $\mathcal{P}_k$ of Borel probability measures on $\mathcal{B}^k.$ Since $|\mathcal{B}^k|=2^k,$ we can identify any $\mu \in \mathcal{P}_k$ with a $2^k$-tuple of real numbers in the closed unit interval that sum to one. We take $\mathcal{P}_k$ to be equipped with the topology that it inherits from being embedded in this way as a closed subset of $\mathbb{R}^{2^k}$ (which we take to be equipped with its standard topology). We call $\mu \in \mathcal{P}^m$ and $\nu \in \mathcal{P}^n$ with $m\leq n$ \emph{consistent} if for each subset $A$ of $\mathcal{B}^m$ we have $\mu(A)=\nu(\pi_{nm}^{-1}(A)).$ \label{itemPk}

\item  \label{itemSubBasic} The space $\mathcal{P}$ of Borel probability measures on $\mathcal{C}$ equipped with the weak topology, which can be characterized as follows.\footnote{For the weak topology on spaces of measures on metric spaces, see  \citep[Chapter II]{parthasarathy2005probability} and  \citep[\S1.2]{Billingsley:1999aa}. For the special case of $\mathcal{P},$ see \citep[Chapter 17]{kechris2012classical} and  \citep[\S 2.5]{reimann2008effectively}. Since we can specify a Borel probability measure on $\mathcal{C}$ by specifying the weight that it assigns to each binary string, by fixing an enumeration of the binary strings we can identify each $\mu \in \mathcal{P}$ with a sequence of numbers in the closed unit interval that sum to one. In this way we identify $\mathcal{P}$ with a closed subset of the Hilbert cube ($=[0,1]^\omega$ equipped with the product topology). The weak topology is the topology that $\mathcal{P}$ inherits from this embedding.}  For each binary string $w$ and each pair of  numbers $p$ and $q$ in the closed unit interval with $p<q,$ let 
\begin{eqnarray*}
S_{w,p,q} := \{ \mu \in \mathcal{P} \,\, : p < \mu(B_w) <q\}.
\end{eqnarray*}
The set of all such $S_{w,p,q}$ forms a sub-basis for the weak topology on $\mathcal{C}$: the open sets of the weak topology are arbitrary unions of finite intersections of these sub-basic sets. Under the weak topology, a sequence $\{\mu_k\}$ of measures in $\mathcal{P}$ converges to  $\mu \in \mathcal{P}$ if and only if $\lim_{k\to \infty} \mu_k(B_w)=\mu(B_w)$  for each $w\in \mathcal{B}^*.$\footnote{Each $B_w$ is a clopen subset of $\mathcal{C}$ and so is a continuity set for any measure in $\mathcal{P}.$ So the Portmanteau Theorem implies that the above condition is necessary for weak convergence. And it is also sufficient, since the $B_w$ form a countable basis for $\mathcal{C}$ closed under finite intersections. See, e.g.,  \citep[Theorems 2.1 and 2.2]{Billingsley:1999aa}.} 

Below, in order to simplify notation, for $\mu$ a measure in $\mathcal{P}$ and $w$ a binary string, we will write $\mu(w)$ in place of $\mu(B_w).$ Using this notation, the Carath\'eodory Extension Theorem tells us that any map $\bar{\nu}: \mathcal{B}^* \to [0,1]$ such that $\bar{\nu}(\varnothing)=1$ and such that $\bar{\nu}(w)=\bar{\nu}(w.0)+\bar{\nu}(w.1)$ for each $w\in \mathcal{B}^*$  induces a unique $\nu \in \mathcal{P}$ such that $\nu(w)=\bar{\nu}(w)$ for all $w\in \mathcal{B}^*$ (see, e.g., \citep[\S 1.9]{nies2009computability}).

As usual, we consider $\nu \in \mathcal{P}$ to be computable if and only if there exists a computable $F: \mathcal{B}^* \times \mathbb{N} \to \mathbb{Q}$ such that $| \nu(w) - F(w,n) | < 2^{-n}$ for all $w\in \mathcal{B}^*$ and $n\in \mathbb{N}.$ We use  $\mathfrak{P}$ to denote subspace of $\mathcal{P}$ consisting of computable measures.

\end{enumerate}

\begin{remark} $\mathcal{B}^*,$ $\mathcal{C},$ $\mathcal{P},$ each of the $\mathcal{B}^k,$ and each of the $\mathcal{P}_k$ are compact, separable  and completely metrizable. In $\mathcal{B},$ the other $\mathcal{B}^n,$ and in $\mathcal{B}^*$ each point is isolated (i.e., for any point, the singleton set containing that point is open). There are no isolated points in $\mathcal{C},$ the $\mathcal{P}_k$ ($k>0$), or $\mathcal{P}.$ \end{remark} 
 
\begin{remark} \label{remCard}
$\mathcal{C}$ and the $\mathcal{P}_k$ ($k>0$) of course have cardinality $\mathfrak{c}$ (the cardinality of the continuum). So does $\mathcal{P}$: since $\mathcal{P}$ is non-empty, compact, and metrizable there is a continuous map from $\mathcal{C}$ onto $\mathcal{P}$; since $\mathcal{P}$ is a non-empty, separable, and completely metrizable space without isolated points, there is an embedding of $\mathcal{C}$ into $\mathcal{P}$ \citep[Theorems 4.18 and 6.2]{kechris2012classical}.   
\end{remark}

\subsection{The Meagre \& the Co-Meagre}

We are going to be interested in making comparisons of size for certain subsets of $\mathcal{C}$ and $\mathcal{P}.$ The most straightforward standard of comparison is cardinality: it natural to say that any uncountable set is incomparably larger than any countable set. 

Below we will see examples where the set of learnable sequences or measures and the set of unlearnable sequences or measures have the same cardinality---but in which it is intuitively natural to say that the unlearnable sequences or measures are incomparably more common than the learnable sequences or measures. 

The intuitive notions of size in play here correspond nicely with the topologists' notions of meagre and co-meagre subsets of a topological space. Recall that a \emph{nowhere dense} subset of a topological space is one whose closure has empty interior---or, equivalently, a subset $A$ of a topological space $X$ is nowhere dense if and only if for any non-empty open set $U \subset X,$ there exists a non-empty open set $U^* \subset U$ with $A \bigcap U^* = \varnothing.$ And recall that a \emph{meagre} subset of a topological space is one that can be written as a countable union of nowhere dense sets while a  \emph{co-meagre} subset is one that is the complement of a meagre set. 

For any topological space $X,$ the class of meagre subsets of $X$ is closed under the operations of taking subsets and taking countable unions. The Baire Category Theorem tells us that in a completely metrizable space, no non-empty open set is meagre. So, in particular, no non-empty completely metrizable space has any subsets that are both meagre and co-meagre.\footnote{If $A\subset X$ were both meagre and co-meagre, then so would be its complement. But then $X$ could be written as a union of two meagre sets---which is impossible if no non-empty open subset of $X$ is meagre.} 

The results just mentioned motivate the standard practice in topology, analysis, and related mathematical fields of considering the elements of a meagre subset of a completely metrizable space to be extremely rare and the  elements of the complement of such a set to be exceedingly common---so that objects that form a co-meagre set are often referred to as being typical. \emph{Illustration}: one says that typical continuous functions on the unit interval are nowhere differentiable because the nowhere differentiable functions form a co-meagre subset of the space of continuous functions under the uniform topology.  

\begin{remark}[The Banach--Mazur Game.] Here is an additional compelling rationale for this practice. Fix a subset $S$ of $\mathcal{C}.$ An infinite two-player game is to be played. In the first round, Player I selects a non-empty binary string $v_1,$ then Player II selects a non-empty binary string $w_1$; and similarly in each subsequent round, Player I selects a non-empty binary string $v_k,$ then Player II selects a non-empty binary string $w_k.$ Player I wins the game if the infinite binary sequence $v_1.w_1.v_2.w_2.\ldots$ is in $S,$ otherwise Player II wins. Intuitively, if Player I has a winning strategy for the the Banach--Mazur game for $S,$ then $S$ must be overwhelmingly large as a subset of $X,$ while if Player II has a winning strategy, then $S$ must be nigh ignorably small as a subset of $X$. 

The intuitive notions of small and large subsets appealed to here correspond precisely to the notions of meagre and co-meagre subsets: Player I has a winning strategy if and only if $S$ is co-meagre in some open subset of $\mathcal{C}$; Player II has a winning strategy if and only if $S$ is meagre as a subset of $\mathcal{C}.$\footnote{Here we have described a special version of the game adapted to $\mathcal{C}.$ A more general version makes sense in any topological space $X$ and we always have the connection between meagreness and winning strategies for Player II; the connection between co-meagreness and winning strategies for Player I requires some additional hypotheses in the general setting.  See \citep{oxtoby1957banach} and \citep[\S\S 8H and 21.C]{kechris2012classical}.} \label{remBM} \end{remark}

\section{Extrapolation} \label{secLearnSeq} 

Think of Nature as having chosen a binary sequence, which is now being revealed to a learning agent one bit at a time. After each new bit is presented, the agent attempts to predict what the next value will be on the basis of the data seen so far. The agent succeeds in this task if from a certain point onwards, the predictions made match reality (almost perfectly). 

\begin{definition}[Extrapolators]
An \emph{extrapolator} is a function $m: \mathcal{B}^* \to \mathcal{B}.$ We denote the set of extrapolators by $\mathcal{E}.$  \end{definition}

\begin{definition}[Extrapolating Machines]
An \emph{extrapolating machine} is a computable extrapol\-ator---i.e., a computable function $m: w\in \mathcal{B}^* \mapsto m(w) \in \mathcal{B}.$ We denote the set of extrapolating machines by $\mathfrak{E}.$  \end{definition}

\begin{definition}[NV-Learning]
Let $m$ be a extrapolator and $\sigma$ a binary sequence. We say that $m$ \emph{NV-learns} $\sigma$ (or that $\sigma$ is \emph{NV-learnable} by $m$) if there is an $N$ such that for all $n>N,$ $m(\sigma[n])=\sigma(n+1).$\footnote{The notion of NV-learning for extrapolating machines is due to B\={a}rzdi\c{n}\v{s} \citep{barzdins1972prognostication}; see also Blum and Blum \citep{blum1975toward}. NV-learning is the subject of an extensive literature---see the canonical surveys on inductive learning, \citep{klette1980research,angluin1983inductive,case1983comparison,odifreddi1992classical, zeugmann2008learning}.\label{fnSurvey}}
\end{definition}

\begin{definition}[Weak NV-Learning] \label{defWeakNV}
We say that $m \in \mathcal{E}$ \emph{weakly NV-learns} $\sigma \in \mathcal{C}$ (or that $\sigma$ is \emph{weakly NV-learnable} by $m$) if:   
\begin{eqnarray*}
\lim_{n\to \infty} \frac{|\{ k\leq n \,\, : \,\,  m(\sigma[k]) = \sigma(k+1) \}|}{n} =1.
\end{eqnarray*}
(i.e., incorrect guesses by $m$ have vanishing limiting relative frequency).
\end{definition}

\begin{remark} \label{remPodniekKinber} Weak NV-learning is a special case (corresponding to $r=1$) of the notion of NV($r$)-learning introduced by Podnieks (in collaboration with Kinber) \citep[pp. 80 f.]{Podnieks:1974aa} in the computable setting: for $r\in (0,1],$ we say that $m\in \mathfrak{E}$ \emph{NV($r$)-learns} $\sigma$ if the correct predictions made by $m$ in processing $\sigma$ have relative frequency at least $r.$\footnote{The notion of NV($r$)-learning appears to have been largely neglected in the subsequent literature (but see \citep{Podnieks:1975aa,Szabo:1987aa}). In particular, it is absent from the surveys cited in fn. \ref{fnSurvey}.}  For expository simplicity, we focus on the special case. But the proofs of the propositions below concerning weak NV-learning can all be adapted to cover NV($r$) learning for any $r \in (0,1]$ (except for Proposition \ref{propSeqWeakEvil}, which requires the restriction $r > 1/2$). \end{remark}


\noindent For $m$ an extrapolator and  $\sigma$ a binary sequence, we say that  according to $m,$  $\sigma(n)$ a \emph{good bit} of $\sigma$ if $m(\sigma[n-1])=\sigma(n)$ and corresponds to a \emph{nasty bit} of $\sigma$ if $m(\sigma[n-1])\neq \sigma(n).$ To say that $m$ NV-learns $\sigma$ is to say that according to $m,$ $\sigma$ eventually consists of nothing but good bits. To say that $m$ weakly NV-learns $\sigma$ is to say that according to $m,$ although $\sigma$ may contain infinitely many nasty bits, these have vanishing limiting relative frequency.  We will consider each of these two criteria of learning in turn. 

\begin{remark}
In the context of computable learners facing computable data streams, one can set the task of   identification (rather than extrapolation): require learners to output a natural number after each bit is revealed, aiming to guess the code number of a program that outputs the data stream they are seeing. We won't discuss this approach extensively here, but will occasionally presume familarity with the literature on identification  problems as surveyed in \citep{klette1980research,angluin1983inductive,case1983comparison,odifreddi1992classical,zeugmann2008learning,jain1999systems}.
\end{remark}

\subsection{NV-Learning} \label{subsecNV}

Officially, the job of an extrapolator is to predict the next bit on the basis of the current data set. But we can also think of an extrapolator $m$ as a means of guessing the entire data sequence on the basis of any initial segment.\footnote{On this point, see  \citep[\S 4.2.1]{angluin1983inductive}.}

\begin{definition}
For $m\in \mathcal{E}$ and $w \in \mathcal{B}^n,$ we use $\sigma_m^w$ to denote the sequence defined as follows: 
\begin{enumerate}[--]
\item For $k=1, \ldots, n,$ $\sigma_m^w(k) = w(k)$ (i.e., $\sigma_m^w[n]=w$).
\item $\sigma_m^w(n+1)=m(w)$;
\item $\sigma_m^w(n+\ell)=m(w.\sigma_m^w(n+1).\ldots. \sigma_m^w(n+\ell-1))$ ($\ell=2,3,\ldots$).
\end{enumerate}
We say that $m$ \emph{guesses} $\sigma_m^w$ on input $w.$
\end{definition} 
    
\noindent Note that if $m\in \mathfrak{E}$ then for any $w\in \mathcal{B}^*,$ $\sigma_m^w \in \mathfrak{C}$: on any input, an extrapolating machine guesses a computable sequence.   

Trivially, there is an equivalence between the sequences NV-learned by an extrapolator and the sequences guessed by it.\footnote{This is related to the deeper fact that $\mathcal{NV}=\mathcal{PEX}$ \citep[Theorem 2.19]{case1983comparison}.}

\begin{proposition} \label{propGuess} Extrapolator $m$ NV-learns sequence $\sigma$ if and only if $\sigma=\sigma_m^w$ for some $w\in \mathcal{B}^*.$  
\proof Suppose that $m$ NV-learns $\sigma.$ Then there is an $n_0$ such that for all $n\geq n_0,$ $m(\sigma[n])=\sigma(n+1).$ So $m$ guesses $\sigma$ on input $w=\sigma[n_0].$ Suppose, on the other hand, there is an $n_0$ such that $m$ guesses $\sigma$ on input $w=\sigma[n_0].$ Then $m$ NV-learns $\sigma,$ since for all $n>n_0,$ $m(\sigma[n])=\sigma(n+1).$
\hfill \qed
\end{proposition}
\noindent So asking that $m$ eventually correctly predict next bits is equivalent to asking that $m$ eventually be able to answer correctly  \emph{all} questions about the data stream. 

\begin{proposition}
For any extrapolator $m,$ the sequences NV-learnable by $m$ form a countably infinite set dense in $\mathcal{C}$ while the sequences not NV-learnable by $m$ form a dense subset of $\mathcal{C}$ of cardinality $\mathfrak{c}.$   
\label{propCountUncount}
\proof On the one hand,  $\mathcal{B}^*$ is a countable set and the preceding proposition tells us that the map $w\in \mathcal{B}^* \mapsto \sigma_m^w$ has as its range the set of sequences NV-learnable by $m.$ So this set is countable. And since for any $w\in \mathcal{B}^*,$ $\sigma_m^w\in B_w$ the set of NV-learnable sequences is dense in $\mathcal{C}$ (and is therefore infinite). On the other hand, each $B_w$ has cardinality $\mathfrak{c}$ but contains only countably many binary sequences NV-learnable by $m.$ 
\hfill \qed
\end{proposition}
\begin{corollary} The set $\{ \sigma \in \mathcal{C} \,\, |\,\, \exists m\in \mathfrak{E} \mbox{ such that }m\mbox{ NV-learns }\sigma\}$ is countable.  
\end{corollary}

As usual, we call a sequence $\{ \sigma_i\}_{i\in \mathbb{N}}$ of elements of $\mathfrak{C}$ \emph{uniformly computable in $i$} if there is a computable $f: \mathbb{N} \times \mathbb{N} \to \mathcal{B}$ such that $f(i,j)=\sigma_i(j),$ for all $i,j\in \mathbb{N}.$

\begin{proposition} \label{propLearnBetter}
(a) Let $m\in \mathcal{E}$  and $S$ be a countable subset of $\mathcal{C}.$ Then there is an $m^* \in \mathcal{E}$ that NV-learns every $\sigma \in S$ as well as everything NV-learned by $m.$ (b) Let $m \in \mathfrak{E}$ and let $S= \{ \sigma_i\}_{i\in \mathbb{N}}$ be a family of elements of $\mathfrak{C}$ uniformly computable in $i.$ Then there is an $m^* \in \mathfrak{E}$ that NV-learns every $\sigma \in S$ as well as everything NV-learned by $m.$
\proof We present the argument for (b)---essentially the same argument works for (a). \\
Define $\tilde{m}\in \mathfrak{E}$ as follows: on input of $w\in \mathcal{B}^n,$ $\tilde{m}$ finds $K= \{ k\in \mathbb{N} \, | \, 1\leq k \leq n, \sigma_k[n]=w\}$; if $K\neq \varnothing,$ then $\tilde{m}(w)=\sigma_\ell(n+1),$ where $\ell$ is the least element of $K$; otherwise, $\tilde{m}(w)=m(w).$ \\
Define $m^*$ as follows: $m^*$ has a counter that keeps tally of how many incorrect prediction have been made in the course of processing a given data stream; in processing input $w\in\mathcal{B}^*,$ $m^*$ simulates $m$ if an even number of incorrect predictions have been made and simulates $\tilde{m}$ if an odd number have been made. \\ 
Clearly, $m^*$ is an extrapolating machine. Suppose that $m^*$ is shown a data stream $\sigma$ that it does not NV-learn. Then $m^*$ must make infinitely many incorrect predictions in processing $\sigma.$ So $\sigma$ cannot be a sequence NV-learned by $m$: any such sequence is guessed by $m$ when it sees sufficiently long initial segments. Similarly, $\sigma$ cannot be any of the $\sigma_k,$ since each of these is guessed by $\tilde{m}$ when it sees sufficiently long initial segments. 
\hfill \qed 
\end{proposition}

\begin{proposition} \label{propEvil}
Let $m$ be an extrapolator and let $S\subset\mathcal{C}$ be the set of sequences that it NV-learns. Then there is an extrapolator $m^\dagger$ such that the set $S^\dagger$ of sequences that it NV-learns is disjoint from $S$---and where $m^\dagger$ is in $\mathfrak{E}$ if $m$ is.  
\proof Define $m^\dagger$ by setting $m^\dagger(w)=1- m(w)$ for each $w\in \mathcal{B}^*.$
\hfill \qed
\end{proposition}

So we have both elements required for the sort of no-free-lunch result we seek. The problem of NV-learning is a formidably difficult one: each (computable) extrapolator fails to NV-learn incomparably more sequences that it NV-learns: the set on which it succeeds is countable (and hence meagre), so the set on which it fails is uncountable (indeed, co-meagre). And there are hard choices to be made: for any (computable) extrapolator, there is another that NV-learns sequences that the first cannot NV-learn. There is no optimal method of extrapolation.

\subsection{Weak NV-Learning} \label{subsecNVw}

If an extrapolator NV-learns a sequence, then it also weakly NV-learns it. But the converse is not true.

\begin{example}
Consider the extrapolating machine $m_1$ that outputs 1 on any input. This machine NV-learns all and only sequences that are eventually all 1's---a countably infinite set. But $m_1$ weakly NV-learns continuum-many sequences. For, let $\hat{\sigma}$ be an arbitrary binary sequence and let $\sigma$ be the sequence defined as follows: for $n=1,2,\ldots,$ if $k=10^n,$ then $\sigma(k)=\hat{\sigma}(n)$; otherwise, $\sigma(k)=1.$ According to $m_1,$ the nasty bits of $\sigma$ have vanishing asymptotic density, so $m_1$ weakly NV-learns $\sigma.$ And there are continuum-many $\hat{\sigma}$ we could use as input for this construction, each determining a distinct sequence weakly NV-learned by $m.$ Note that there is no input on which $m_1$ guesses a sequence that contains infinitely many 0's, although it weakly NV-learns uncountably many sequences with this feature. Note also that although $m_1$ is computable, it weakly NV-learns uncountably many uncomputable sequences and weakly NV-learns sequences of arbitrary Turing degree. \label{exM0}
 \end{example}

\begin{proposition} \label{propWeakDense}
Each extrapolator weakly NV-learns a dense set of sequences of cardinality $\mathfrak{c}$ and fails to weakly NV-learn a dense set of sequences of cardinality $\mathfrak{c}.$ 
\proof  Let $m$ be an extrapolator, $w$ an $n$-bit binary string, and $\hat{\sigma}$ an arbitrary sequence. We construct sequences $\sigma^*$ and $\sigma^\dagger$ as follows:  
\begin{enumerate}[--]
\item For $k=1, \ldots, n,$ $\sigma^*(k)=\sigma^\dagger(k)=w(k).$ 
\item For $k= n + 10^\ell$ ($\ell=1,2,\ldots$), $\sigma^*(k)=\sigma^\dagger(k)=\hat{\sigma}(\ell).$ 
\item For all other $k,$ $\sigma^*(k)=m(\sigma^*(1).\sigma^*(2).\ldots.\sigma^*(k-1))$ and  $\sigma^\dagger(k)=1- \sigma^*(k).$
\end{enumerate}
According to $m,$ any nasty (good) bits in $\sigma^*$ ($\sigma^\dagger$) occur with indices of the form $n + 10^\ell.$ So $m$ weakly NV-learns $\sigma^*$ and fails to weakly NV-learn $\sigma^\dagger.$ By varying $w,$ we obtain weakly NV-learnable and not weakly NV-learnable sequences in each basic open set of $\mathcal{C}.$ And by varying $\hat{\sigma}$ we obtain continuum-many sequences of each type. \hfill \qed
\end{proposition}

So for any extrapolator, there are continuum-many sequences that it can weakly NV-learn and continuum-many sequences that it cannot weakly NV-learn. But, intuitively, there is a sense in which it is much more difficult to construct a sequence weakly NV-learnable by a given extrapolator than it is to construct a sequence that is not weakly NV-learnable by that extrapolator. Consider again the extrapolator $m_1$ that outputs 1 on any input. In order to construct a sequence that this extrapolator weakly NV-learns, you begin with the all 1's sequence, then sprinkle in some 0's, subject to the constraint that the set of indices of the slots containing 0's has vanishing asymptotic density in $\mathbb{N}.$ In order to construct a sequence that this extrapolator \emph{can't} weakly NV-learn, you begin with the all 1's sequence and sprinkle in as many 0's as you like, just being careful to make sure that the set of indices of the slots containing 0's doesn't have vanishing asymptotic density. The latter task, is intuitively, easier: e.g., because there are a lot more densities not equal to zero than equal to zero.  This intuition is borne out by the following result. 

\begin{proposition}
Let $m$ be any extrapolator. The sequences weakly NV-learnable by $m$ form a meagre subset of $\mathcal{C}.$ \label{propSeqWeakMeagre}
\proof 
Let us say that binary string $w$ is \emph{wicked} according to $m$ if at least half of the bits of $w$ are nasty according to $m.$ For each $n \in \mathbb{N},$ let $A_n$ be the set of sequences that do not have at least $n$ initial segments that are wicked according to $m.$ \\
We claim that each $A_n$ is nowhere dense. To establish this, it suffices to show that for any binary string $w,$ there is another, $w^*,$ depending on $n$ and $w,$ such that $w^*$ extends $w$ and $B_{w^*}\bigcap A_n = \varnothing.$ To this end, let $w$ be a string and let $w^*$ be the result of extending $w$ by $|w|$ bits that are nasty according to $m,$ then tacking on $n$ more nasty bits. Every sequence in $B_{w^*}$ then has at least $n$ initial segments that are wicked according to $m.$ \\
So $A:= \bigcup_{n=1}^\infty A_n$ is a meagre subset of $\mathcal{C}.$ And any sequence $\sigma$ weakly NV-learnable by $m$ must be in  $A$---for otherwise, $\sigma$ would have the feature that for each $k,$ it contained at least $k$ initial segments wicked according to $m,$ which would mean that the asymptotic density of nasty bits in $\sigma$ could not vanish. So the set of sequences weakly NV-learnable by $m,$ being a subset of a meagre set, is meagre. 
\hfill \qed
\end{proposition}
\begin{corollary} The set $\{ \sigma \in \mathcal{C} \,\, |\,\, \exists m\in \mathfrak{E} \mbox{ such that }m\mbox{ weakly NV-learns }\sigma\}$ is meagre in $\mathcal{C}.$\footnote{Remark \ref{remCoarse} below will show that this strengthens the observation of Jockusch and Schupp \citep[p. 438]{jockusch2012generic} that the set of coarsely computable sequences is meagre in $\mathcal{C}.$}
\end{corollary}

So the problem of weakly NV-learning sequences is formidably difficult. And difficult choices must be made in the face of this intractability---there can be no optimal extrapolator for weak NV-learning.

We have the following better-but-no-best result.\footnote{Thanks here to Tom Sterkenburg and to an anonymous referee for helpful suggestions.}

\begin{proposition} 
(a) Let $m \in \mathcal{E}$ and let $S$ be a countable subset of $\mathcal{C}.$ Then there is an $m^* \in \mathcal{E}$ that NV-learns every $\sigma \in S$ and also weakly NV-learns everything that $m$ does. (b) Let $m \in \mathfrak{E}$ and let $S=\{ \sigma_i\}_{i\in \mathbb{N}}$ be a family of elements of $\mathfrak{C}$ that is uniformly computable in $i.$ Then there is an $m^* \in \mathfrak{E}$ that NV-learns every $\sigma \in S$ and also weakly NV-learns everything that $m$ does. \label{propWeakLearnBetter}
\proof We present the argument for (b)---essentially the same argument works for (a). \\ 
Define $m^*$ as follows: on input of $w\in \mathcal{B}^n,$ $m^*$ finds $K= \{ k\in \mathbb{N} \, | \, 1\leq k \leq \log_2 n, \sigma_k[n]=w\}$; if $K\neq \varnothing,$ then $m^*(w)=\sigma_\ell(n+1),$ where $\ell$ is the least element of $K$; otherwise, $m^*(w)=m(w).$  Clearly, $m^*$ is an extrapolating machine and NV-learns each $\sigma_k$ ($m^*$ guesses $\sigma_k$ whenever shown sufficiently long initial segments).  And if  $\sigma \in \mathcal{C}$ is weakly NV-learned by $m$ then it is also weakly NV-learned by $m^*$: in processing the first $2^n$ bits of $\sigma,$ $m^*$ can disagree with $m$ at most $n$ times; so the asymptotic density of bits on which $m^*$ and $m$ disagree in processing $\sigma$ is zero. \hfill \qed 
\end{proposition}

\noindent We also have the usual sort of evil-twin result. 

\begin{proposition} \label{propSeqWeakEvil}
Let $m$ be an extrapolator and let $S\subset\mathcal{C}$ be the set of sequences that it weakly NV-learns. Then there is an extrapolator $m^\dagger$ such that the set $S^\dagger$ of sequences that it weakly NV-learns is disjoint from $S$---and where $m^\dagger$ is in $\mathfrak{E}$ of $m$ is. 
\proof Define $m^\dagger$ by setting $m^\dagger(w)=1- m(w)$ for each $w\in \mathcal{B}^*.$ According to either $m$ or $m^\dagger,$ in any sequence that the other weakly NV-learns, the good bits have asymptotic density zero.  
\hfill \qed
\end{proposition}

\subsection{Extrapolation of Computable Sequences} \label{subsecCompSeq}

While it is  plausible that every method of learning implementable by a natural or artificial learning agent is computable, the data streams that our agents face may or may not be computable.\footnote{Unless, that is, physical reality itself is fundamentally computational in nature---for a range of views of this topic, see the papers collected in \citep{zenil2013computable}.} Still, there are many settings in which we can be confident that our agents face computable data streams. So let us specialize to the setting  in which computable extrapolators attempt to (weakly) NV-learn computable sequences and see how the landscape surveyed above is transformed.

As usual, for $m\in \mathfrak{E},$  we denote by $\mathcal{NV}(m)$ the set of computable sequences that are NV-learned by $m.$  We use $\mathcal{NV}$ to denote:
\begin{eqnarray*}
\{ S \subset \mathfrak{C} \,\, | \,\, \exists m \in \mathfrak{E} \mbox{ with } S \subseteq \mathcal{NV}(m) \}.
\end{eqnarray*}
We likewise use $\mathcal{NV}^w(m)$ to denote the set of computable sequences weakly NV-learned by an extrapolating machine $m$ and use $\mathcal{NV}^w$ to denote:
\begin{eqnarray*}
\{ S \subset \mathfrak{C} \,\, | \,\,  \exists m \in \mathfrak{E} \mbox{ with } S \subseteq \mathcal{NV}^w(m)\}.
\end{eqnarray*}

\begin{proposition}[Podnieks \citep{Podnieks:1974aa}] \label{exBC}
$\mathcal{NV}$ is a proper subset of $\mathcal{NV}^w.$
\proof Clearly $\mathcal{NV} \subseteq \mathcal{NV}^w.$ We give an example of a set in $\mathcal{NV}^w - \mathcal{NV}.$\\
Consider again the extrapolating machine $m_1$ of Example \ref{exM0} above that outputs 1 on every input. Let $U= \mathcal{NV}^w(m_1),$ the set of computable binary sequences in which 0's have vanishing asymptotic density. We are going to show that $U$ is not in $\mathcal{NV}.$  \\
Suppose that there is an  extrapolating machine $m$ that NV-learns each sequence in $U.$ Notice that for any $w\in \mathcal{B}^*,$ the sequence $w.1^\omega$ is in $U$---so for sufficiently large $\ell \in \mathbb{N},$ we must have  $m(w.1^\ell)=1.$ Let $\sigma \in \mathcal{C}$ be the sequence of the form $1^{n_1}.0.1^{n_2}.0.1^{n_3}.0\ldots$ where each $n_j$ is chosen to be the smallest $n$ larger than $2^j$ such that $m(1^{n_1}.0.1^{n_2}.0.\ldots.1^{n_{j-1}}.0.1^n)=1.$ Clearly, $m$ does not NV-learn $\sigma$ ($\sigma$ contains infinitely many 0's, each of which $m$ predicts will be a 1). But $\sigma \in U$: since $m$ is computable, so is $\sigma$; and by construction, 0's occur with vanishing asymptotic density in $\sigma.$ This contradicts our assumption that $m$ NV-learns every $\sigma \in U.$ \hfill \qed
\end{proposition}

\begin{remark}
We mention two of the most fundamental variations on $\mathcal{NV}.$ A \emph{partial extrapolating machine} is a partial computable function $m: \mathcal{B}^* \to \mathcal{B}.$  Following B\={a}rzdi\c{n}\v{s} \citep{barzdins1972prediction}, we say that a partial extrapolating machine $m$ \emph{NV\hspace{1pt}$^\prime$-extrapolates} $\sigma\in \mathfrak{C}$ if: (i) $m(\sigma[k])$ is defined for all $k \in \mathbb{N}$; and (ii) $\exists N \in \mathbb{N}$ such that for all $n>N,$ $m(\sigma[n])=\sigma(n+1).$ We write $\mathcal{NV}\hspace{1pt}^\prime$ for the set of $S \subset \mathfrak{C}$ such that there is a partial extrapolating machine that NV\hspace{1pt}$^\prime$-extrapolates each $\sigma \in S.$ \\
Following Podnieks \citep{Podnieks:1974aa}, we say that a partial extrapolating machine $m$ \emph{NV$\hspace{1pt}^{\prime\prime}$-extrapolates} $\sigma\in \mathfrak{C}$ if:  $\exists N \in \mathbb{N}$ such that for all $n>N,$ $m(\sigma[n])$ is defined and equal to $\sigma(n+1).$ We write $\mathcal{NV}\hspace{1pt}^{\prime\prime}$ for the set of $S \subset \mathfrak{C}$ such that there is a partial extrapolating machine that NV$\hspace{1pt}^{\prime\prime}$-extrapolates each $\sigma \in S.$\\
Obviously, $\mathcal{NV} \subseteq \mathcal{NV}\hspace{1pt}^\prime \subseteq \mathcal{NV}\hspace{1pt}^{\prime\prime}.$ In fact, $\mathcal{NV} \subset \mathcal{NV}\hspace{1pt}^\prime \subset \mathcal{NV}\hspace{1pt}^{\prime\prime}.$\footnote{That $\mathcal{NV}\subset \mathcal{NV}\hspace{1pt}^\prime$ is due to B\={a}rzdi\c{n}\v{s} \citep{barzdins1972prediction}; that $\mathcal{NV}\hspace{1pt}^\prime \subset \mathcal{NV}\hspace{1pt}^{\prime\prime}$ is due to Podnieks \citep{Podnieks:1974aa}. See \citep[Corollary 2.29, Corollary 2.31, Theorem 3.1, and Theorem 3.5]{case1983comparison}.} The proof of Proposition \ref{exBC} above carries over essentially unchanged (except that dove-tailing is required) to show that $\mathcal{NV}^w$ is not contained in $\mathcal{NV}\hspace{1pt}^{\prime\prime}$ (let alone in $\mathcal{NV}\hspace{1pt}^\prime$). We will see below in Remark \ref{remFortnow} that $\mathcal{NV}^w$ does not contain $\mathcal{NV}\hspace{1pt}^\prime$ (let alone $\mathcal{NV}\hspace{1pt}^{\prime\prime}$).   
\end{remark}

\begin{remark}
Having introduced, for $r\in (0,1],$ the notion of NV($r$) learning (see Remark \ref{remPodniekKinber} above),  Podnieks (in collaboration with Kinber) \citep{Podnieks:1974aa,Podnieks:1975aa} introduces the classes $\mathcal{NV}(r),$ $\mathcal{NV}\hspace{1pt}^\prime(r),$ and $\mathcal{NV}\hspace{1pt}^{\prime\prime}(r)$ in the obvious way. Podnieks goes on  \citep{Podnieks:1974aa,Podnieks:1975aa} to establish a number of facts about the containment relations involving these classes.  For present purposes, the most notable are: if $r_1 < r_2,$ then $\mathcal{NV}(r_2)$ is a proper subset of $\mathcal{NV}(r_1)$; and $\mathfrak{C} \in \mathcal{NV}\hspace{1pt}^{\prime\prime}(1)$ (from which it follows that $\mathcal{NV}\hspace{1pt}^{\prime\prime}$ is properly contained in $\mathcal{NV}\hspace{1pt}^{\prime\prime}(1)$). \\
It is also possible to introduce up-to-exceptions-of-density-$r$ variants of standard identification classes. Podnieks \citep{Podnieks:1974aa} generalizes $\mathcal{BC}$ by considering learners who aim to output at each time (modulo permitted exceptions) a code number for a program that generates the data stream they see. Pitt \citep{pitt1989probabilistic} generalizes $\mathcal{EX}$ by considering learners who aim to converge (modulo permitted exceptions) to a single such code number.\footnote{Both these notions turn out to be closely related by identification by probabilistic learners to identification by teams of learners \citep{pitt1989probabilistic,zeugmann2008learning}. We do not have the same close association between the corresponding notions in the extrapolation case} Royer \citep{royer1986inductive,jain1999systems} generalizes $\mathcal{EX}$ in a different direction, considering learners who aim to converge to a single code number that (modulo permitted exceptions) outputs the data stream they are seeing. Jain \citep{jain1996program} investigates the problem of identification of a function from data streams that are accurate only modulo errors of given asymptotic density.
\label{remPitt}
\end{remark}

Via Propositions \ref{propGuess} and \ref{propCountUncount}, we know that every extrapolating machine NV-learns a countably infinite subset of $\mathfrak{C}.$ But there there can be no best extrapolating machine: Propositions \ref{propEvil} and \ref{propSeqWeakEvil} tell us each extrapolation machine has an evil twin that (weakly) NV-learns a disjoint set of computable sequences; and Propositions \ref{propLearnBetter}(b) and \ref{propWeakLearnBetter}(b) tell us that each extrapolating machine is dominated by another that (weakly) NV-learns everything it can while also NV-learning every member of a uniformly computable family of elements of $\mathfrak{C}.$
 
In this setting, what comparative judgements can we make about the sets of computable sequences that a given extrapolating machine (weakly) NV-learns and doesn't (weakly) NV-learn? 

\begin{proposition}
For any  $m\in \mathfrak{E},$ following are dense subsets of $\mathcal{C}$:
\begin{enumerate}[(a)]
\item the set of computable sequences NV-learnable by $m$;
\item the set of computable sequences not NV-learnable by $m$;
\item the set of computable sequences weakly NV-learnable by $m$;
\item the set of computable sequences not weakly NV-learnable by $m.$
\end{enumerate}
Straightforward adaptations of the proofs of Propositions \ref{propCountUncount} and  \ref{propWeakDense} yield that (a) and (d) are dense. And (b) and (c) are super-sets of (d) and of (a), respectively. 
\hfill \qed
\end{proposition}

It follows that the set of computable sequences (weakly) NV-learned by an extrapolating machine $m$ and the set of computable sequences not (weakly) NV-learned by an extrapolating machine $m$ are both countably infinite subsets of $\mathcal{C}$---so we have parity at the level of cardinality. A classical result implies that we also have parity at the level of topology. 

\begin{proposition}[Sierpi\'nski] Any two countable dense subsets of $\mathcal{C}$ are homeomorphic.\label{propSierpinski}
\proof See, e.g.,  \citep[Chapter 17]{dasgupta2014set}. 
\hfill \qed
\end{proposition}

So we can say: for any computable method of extrapolating computable sequences, failure and success are equally common---and difficult choices must be made in selecting a computable method of extrapolation, since no method dominates all its rivals in its range of success. 

But, intuitively, we ought to be able to say something stronger. After all, B\={a}rzdi\c{n}\v{s} \citep{barzdins1972prediction} showed that if $S$ is a set of computable functions, then following are equivalent: (i) each member of $S$ is NV-learnable; (ii) $S$ is a subclass of a computably enumerable set of computable functions; (iii)  $S$ is a subclass of an abstract complexity class.\footnote{See also Blum and Blum \citep[p. 127]{blum1975toward}, who attribute the complexity-theoretic condition independently to Adleman. As Blum and Blum remark, this result shows ``in essence, that the extrapolable sequences are the ones that can be computed rapidly.''} So only very special subsets of $\mathfrak{C}$ are NV-learnable---which means that generic subsets should not be in $\mathcal{NV}.$ 

Indeed, there is a natural hybrid computational-topological notion of that underwrites the conclusion that failure is incomparably more common than success for computable extrapolation of computable sequences. Mehlhorn \citep{mehlhorn1973size} introduced the important notion of an effectively meagre subset of the set of computable functions. We specialize this apparatus to   $\mathfrak{C}.$ 

By way of motivation, note that in any topological space $X$ with basis of open sets $\mathcal{W},$ a subset $A$ is nowhere dense if and only if for every non-empty $U\in \mathcal{W}$ there is a non-empty $U^* \in \mathcal{W}$ with $U^* \subset U$ such that $A \bigcap U^* = \varnothing.$  So a subset $A\subset \mathcal{C}$ is nowhere dense if and only if there is a function $f : \mathcal{B}^* \to \mathcal{B}^*$ such that for each binary string $w$: (i)  $f(w)$ extends $w$; and (ii) $A \bigcap B_{f(w)} = \varnothing.$ And $A \subset \mathcal{C}$ is meagre if and only if there is a function $F: \mathbb{N} \times \mathcal{B}^* \to \mathcal{B}^*$ such that: (i) for each $n\in \mathbb{N}$ there is an $A_n \subset \mathcal{C}$ such that $f_n = F(n, \cdot)$ is a witness to the fact that $A_n$ is nowhere dense in $\mathcal{C}$; and (ii) $A = \bigcup_{n\in \mathbb{N}} A_n.$ 

\begin{definition}[Mehlhorn \citep{mehlhorn1973size}] Let  $A$ be a subset of $\mathfrak{C}$ and let $f: \mathcal{B}^* \to \mathcal{B}^*$ be a computable function. Then $A$ is \emph{effectively nowhere dense via} $f$ if for each $w \in \mathcal{B}^*$: 
\begin{enumerate}[i)]
\item $f(w)$ extends $w$;
\item $A \bigcap B_{f(w)} = \varnothing.$
\end{enumerate}
\end{definition} 

\begin{definition}[Mehlhorn \citep{mehlhorn1973size}] A subset $A$ of $\mathfrak{C}$ is \emph{effectively meagre} if there is a computable function $F: \mathbb{N} \times \mathcal{B}^* \to \mathcal{B}^*$ such that:
\begin{enumerate}[i)]
\item for each $n \in \mathbb{N},$ there is an $A_n \subset \mathfrak{C}$ such that $A_n$ is effectively nowhere dense via $f_n= F(n, \cdot)$; 
\item $A \bigcup_{n\in \mathbb{N}} A_n.$
\end{enumerate}
The complement in $\mathfrak{C}$ of an effectively meagre subset of $\mathfrak{C}$ is called \emph{effectively co-meagre}.
\end{definition} 

\begin{proposition}[Mehlhorn \citep{mehlhorn1973size}] \label{propMehl1}
The family of effectively meagre subsets of $\mathfrak{C}$ is closed under the following operations:
\begin{enumerate}[a)]
\item taking subsets;
\item taking finite unions;
\item taking effective unions.
\end{enumerate}
\proof The first claim is immediate from the definition and the second follows from the third. So suppose that that $M$ is a subset of $\mathfrak{C}$ such that there exist a computable $H:\mathbb{N} \times \mathbb{N} \times \mathcal{B}^*$ and a decomposition $M=\bigcup N_i,$ such that for each $k \in \mathbb{N},$ $H(k,\cdot, \cdot)$ is a witness to the fact that $N_k$ is effectively meagre. There exists, then, for each $i \in \mathbb{N},$  a decomposition $N_i= \bigcup N_{ij}$ such that each $N_{ij}$ is effectively nowhere dense in virtue of $H(i,j,\cdot).$ Fix a computable bijection $\pi: \mathbb{N} \times \mathbb{N} \to \mathbb{N}$ and let $p_1$ and $p_2$ be the computable components of the inverse of $\pi$ (so that $\pi (p_1(k),p_2(k))=k$ for all $k\in \mathbb{N}$). Set $M_k:=N_{p_1(k),p_2(k)}$ and for each $w \in \mathcal{B}^*,$ set $F(k,t):=H(p_1(k),p_2(k),w).$ Then $F: \mathbb{N} \times \mathcal{B}^*$ is computable, $M= \bigcup M_k,$ and each $M_k$ is effectively nowhere dense in virtue of $F(k,\cdot).$ So $M$ is effectively meagre. 
 \hfill \qed
\end{proposition}

\noindent Crucially, the set of effectively meagre subsets of $\mathfrak{C}$ is not closed under arbitrary countable unions due to an effective analog of the Baire Category Theorem. 

\begin{proposition}[Mehlhorn \citep{mehlhorn1973size}] \label{propMehl2}Let $w$ be a binary string. Then $B_w \bigcap \mathfrak{C}$ is not effectively meagre. 
\proof  Let $M=\bigcup M_k$ be an effectively meagre set with witness $F: \mathbb{N} \times \mathcal{B}^*.$
We construct strings $w_0,$ $w_1,$ \ldots\ inductively: $w_0:=w$; and $w_{k+1} = F(k, w_k).0.$ By construction, each $w_k$ is a proper initial segment of $w_{k+1}.$ Let $\sigma = \lim_{n\to \infty} w_n.$ Then $\sigma\in B_w \bigcap \mathfrak{C}.$ But for each $k,$ $\sigma \notin M_k$ (since $\sigma$ begins with $w_{k+1}$), so $\sigma \notin M.$  
\hfill \qed
\end{proposition}

In light of these results, it is natural to think of elements of effectively meagre subsets of $\mathfrak{C}$ as being incomparably less common than elements of effectively co-meagre subsets of $\mathfrak{C},$ even when the meagre and co-meagre sets in question are both dense as subsets of $\mathcal{C}.$ 

\begin{remark} \label{remEffectiveBM} A further reason (due to Lisagor \citep{lisagor1981banach}) for this standard practice: a subset $S$ of $\mathfrak{C}$ is effectively meagre if and only if when the Banach--Mazur game (described in Remark \ref{remBM} above) is played for $S,$ Player II has a winning strategy that is computable. Another reason (due, again, to Mehlhorn \citep{mehlhorn1973size}): each abstract complexity class is effectively meagre as a subset of the family of computable functions.
\label{remCalude}
\end{remark}

\begin{example}[Self-Describing Sequences]  \label{exFortnow} Fix an enumeration $M_1,$ $M_2,$ \ldots\ of the Turing machines, with associated acceptable programming system $\phi_1,$ $\phi_2,$ \ldots\ (so that $\phi_k$ is the partial computable function computed by $M_k$).  Following Blum and Blum \citep{blum1975toward}, we call a sequence $\sigma \in \mathfrak{C}$ \emph{self-describing} if $\sigma$ has an initial segment of the form $1^k0$ and is computed by $M_k.$ As  Blum and Blum  note, the set $S_1$ of self-describing sequences is non-trivial: it follows from the Recursion Theorem that each computable binary sequence is a finite variant of a self-describing sequence---so there are arbitrarily complex sequences in $S_1.$ \\ 
Fortnow \emph{et al.} \citep{fortnow1998relative} observe that $S_1$ is not effectively meagre. For, consider any computable strategy $\beta: \mathcal{B}^n \to \mathcal{B}^n$ that Player II could use to play the Banach--Mazur game for $S_1.$ For each $k \in \mathbb{N},$ let $\alpha_k$ be the following strategy that Player I might adopt: on the first turn, play $1^k0$; on all subsequent turns, play 0. The assumption that Player I plays strategy $\alpha_k$ and Player II plays strategy $\beta$ determines a unique sequence $\sigma_k\in \mathcal{C}.$ The map $F: (k,\ell)\in \mathbb{N}^2 \mapsto \sigma_k(\ell) \in \mathcal{B}^*$ is computable. So by the Recursion Theorem, there is a $k_0\in \mathbb{N}$ such that $\sigma_{k_0}$ is computed by $M_{k_0}.$ That is: there exists a strategy (namely, $\alpha_{k_0}$) via which Player I can defeat $\beta.$ So $S_1$ is not in effectively meagre. 
\end{example}

\begin{proposition} \label{propCompWeakMeagre}
Let $m$ be an extrapolating machine. $\mathcal{NV}^w(m)$ (the set of computable sequences weakly NV-learnable by $m$) is an effectively meagre subset of $\mathfrak{C}.$
\proof A straightforward adaptation of the proof of Proposition \ref{propSeqWeakMeagre}, appealing to the fact that  when $m$ is computable, the map $(n,w) \mapsto w^*\notin A_n$ used there is computable. 
\hfill \qed
\end{proposition}
\begin{corollary}[Fortnow \emph{et al.} \citep{fortnow1998relative}]
Let $m$ be an extrapolating machine. $\mathcal{NV}(m)$ (the set of computable sequences NV-learnable by $m$) is an effectively meagre subset of $\mathfrak{C}.$ \label{corFortnow}
\end{corollary}
So there is a natural sense in which, for any computable extrapolator $m,$ among computable sequences, those (weakly) NV-learnable by $m$ are incomparably less common than those not (weakly) NV-learnable by $m.$ The problem of (weakly) NV-learning computable sequences is formidably difficult. 
\begin{corollary}
No extrapolating machine can (weakly) NV-learn each self-describing sequence. 
\end{corollary}

\begin{remark} \label{remFortnow}
It is illuminating to situate these results with respect to a couple of results that Fortnow \emph{et al.} \citep{fortnow1998relative} establish concerning the identification problem for binary sequences.
\begin{enumerate}[i)]
\item They show via an effective Banach--Mazur argument,  that any $S \in \mathcal{PEX}$ is effectively meagre. 
\item They observe that the set of self-describing functions (see Example \ref{exFortnow} above) is in $\mathcal{EX}_0.$\footnote{Consider a learner who is silent until a data set of the form $1^k0$ is seen  and who from then on assumes that the data stream is being generated by $M_k.$} So any identification class that contains $\mathcal{EX}_0$ has members that are not effectively meagre. 
\end{enumerate}
They remark: ``Since virtually every inference class is either a subset of $\mathcal{PEX}$ or a superset of $\mathcal{EX}_0$ the results here settle virtually all open questions that could be raised'' \citep[p. 145]{fortnow1998relative}. \\
Contact can be made with the present approach by recalling that $\mathcal{PEX}=\mathcal{NV}$ and that $\mathcal{EX}_0 \subset \mathcal{NV}\hspace{1pt}^\prime.$\footnote{See Case and Smith \citep{case1983comparison}: that $\mathcal{PEX}=\mathcal{NV}$ is their Theorem 2.19 (attributed to private communications from van Leeuwen and B\={a}rzdi\c{n}\v{s}); that $\mathcal{EX}_0 \subset \mathcal{NV}\hspace{1pt}^\prime$ is their Theorem 2.28.\label{fnLeeuwen}} So the first result of Fortnow \emph{et al.} \citep{fortnow1998relative} noted above is our Corollary \ref{corFortnow}: the set of sequences NV-learnable by an extrapolating machine is effectively meagre. And since every set in $\mathcal{NV}^w$ is effectively meagre, neither $\mathcal{NV}\hspace{1pt}^\prime$ nor $\mathcal{NV}\hspace{1pt}^{\prime\prime}$ is a subset of $\mathcal{NV}^w.$ In $\mathcal{NV}^w$ we have an example of a natural inference class  that is neither a subset of $\mathcal{PEX}$ nor a superset of $\mathcal{EX}_0.$ \\
In the present setting, in which computable extrapolators (i.e., extrapolating machines) attempt to learn computable sequences, we find that generalizing our basic model by allowing merely partially defined extrapolating machines allows us to crash through a size barrier in a way that loosening our criterion of success by allowing infinitely many errors in the sense of weak NV-learning does not---since every set in $\mathcal{NV}$ or in $\mathcal{NV}^w$ is effectively meagre, whereas this is not the case for every set in $\mathcal{NV}\hspace{1pt}^\prime$ or $\mathcal{NV}\hspace{1pt}^{\prime\prime}.$ This is the reverse of what we find if we challenge (possibly computable) extrapolators to NV-learn arbitrary sequences. In that setting, in the basic model every learner masters only countably many sequences. And this is unchanged if we countenance merely partially defined learners.\footnote{Thanks to an anonymous referee for pointing this out.} But if we loosen our criterion of success to weak NV-learnability we crash through a cardinality barrier, as each learner weakly NV-learns uncountably many sequences.\end{remark}

\begin{remark}[Coarse Computability]  \label{remCoarse}
Jockusch and Schupp \citep[p. 472]{jockusch2012generic} remark that ``In recent years, there has been a general realization that worst-case complexity measures, such as $P,$ $NP,$ exponential time, and just being computable, often do not give a good overall picture of the difficulty of a problem.'' As an example, they observe that although there exist finitely presented groups with unsolvable word problems, in every such group the words expressing the identity have vanishing asymptotic density, when words are enumerated in lexicographic order. So the linear-time algorithm that on the input of any word guesses that that word does not express the identity would make a negligible set of errors if fed all words in lexicographic order. If we demand perfection, then the word problem is impossibly hard---but if we can live with making mistakes a negligible fraction of the time, it is as easy as could be. This motivates Jockusch and Schupp to introduce a generalization of the notion of computability: a sequence is \emph{coarsely computable} if it differs from some computable sequence in a set of bits of vanishing asymptotic density. 

Coarse computability implies computable weak NV-learnability: if $\sigma \in \mathcal{C}$ differs from $\sigma^* \in \mathfrak{C}$ only in bits of vanishing asymptotic density, then the extrapolating machine that assumes it is being shown $\sigma^*$ on any input weakly NV-learns $\sigma.$

But weak NV-learnability does not imply coarse computability. Let $\sigma_0$ be uncomputable. Construct a sequence $\sigma_1$ as follows: begin with two copies of the first bit of $\sigma_0,$ followed by four copies of the second bit of $\sigma_0,$ \ldots\ followed by $2^k$ copies of the $k$th bit of $\sigma_0,$ \ldots .  Suppose that $\sigma_1$ is coarsely computable. Then there must be a computable sequence $\sigma_2$ that differs from $\sigma_1$ only in a set of bits of vanishing asymptotic density. Define a new sequence $\sigma_3$ as follows: make the first bit of $\sigma_3$ a 0 if at least one of the first two bits of $\sigma_2$ is a 0, otherwise make it a 1; make the second bit of $\sigma_3$ a 0 if at least two of the next four bits of $\sigma_2$ are 0, otherwise make it a 1; \ldots; make the $k$th bit of $\sigma_3$ a 0 if at least $2^{k-1}$ of the next $2^k$ bits of $\sigma_2$ are 0, otherwise make it a 1; \ldots. Since $\sigma_2$ is computable (by assumption), so is $\sigma_3.$ But $\sigma_3$ is a finite variant of $\sigma_0$ and so must be uncomputable. So there can be no such $\sigma_2$: $\sigma_1$ is not coarsely computable. But $\sigma_1$ is weakly NV-learned by the extrapolating machine that predicts the first bit will be a 1 then subsequently predicts that each bit will be the same as the last bit seen. 
\end{remark}

\section{Forecasting} \label{secLearnMeas}

So far we have set our learners the problem of recognizing which binary sequence is being revealed in the data stream---where such recognition consists in becoming good at predicting future bits. In effect, we have been picturing that in generating new bits, Nature simply consults a lookup table chosen in advance and that the learner's job is to attempt to guess which of the possible such tables is being used (or, in the case of weak learning, to attempt to come close to guessing the right table, in a certain sense). 

We might instead picture a different sort of procedure. Suppose that what Nature has chosen in advance is not a sequence but, rather, a measure $\lambda \in \mathcal{P}$ (i.e., a Borel probability measure on $\mathcal{C}$) and that the learner's data stream is randomly sampled from $\lambda.$ So we now picture Nature as being equipped with a complete set of biased coins and an instruction manual that says which coin to toss to generate the next bit, given the bits that have been generated so far. To mention just some of the tamest possibilities: Nature may have chosen a Bernoulli measure, in which case the instruction will be to use the same  coin to generate each new bit; or Nature may have chosen a measure corresponding to a Markov chain, in which case the coin chosen to generate a new bit will depend only on some fixed finite number of immediately preceding bits; or Nature could  have chosen a delta-function measure concentrated on a single sequence, in which case only a maximally biased coin will ever be used.

\begin{definition}[Sources]
A \emph{source} is a Borel probability measure on  $\mathcal{C}.$ 
\end{definition} 

In what follows, we will think of Nature as having chosen a source $\lambda \in \mathcal{P}$  from which our learner's data stream is sampled. Recall for $w\in \mathcal{B}^*,$ we write $\lambda(w)$ in place of $\lambda(B_w).$ Similarly, for $s=0,1$ and $w\in \mathcal{B}^n,$ we will write $\lambda(s \, | \, w)$ for the conditional probability $\lambda$ gives for the $(n+1)$st bit to be $s$ given that the first $n$ bits were given by $w.$

How should a learner proceed in the setting where the data stream is given by a probabilistic source? In the setting of Section \ref{secLearnSeq}, where we were thinking of new bits as being generated by a deterministic process, we asked learners to choose an extrapolator that would allow them to definitively predict at each stage what the next bit would be, given the data seen so far. That approach would be suboptimal in the present setting: if Nature is using the fair coin measure (the Bernoulli measure of bias .5) to generate the data stream, then (with probability one) no extrapolator will do better (or worse) than random in its predictions of the next bit---but the fact that Nature is using this procedure seems like a paradigm example of the sort of thing that we ought to be able to learn by looking at data. Such learning will be possible if  we ask agents to choose a forecasting procedure that allows them to issue a forecast probability before each bit is revealed, rather than choosing an extrapolator that at each stage issues definitive predictions regarding the next bit. 

A natural way to encode such a strategy for learning would be via a \emph{confirmation function}: a map $\tilde{\mu}: \mathcal{B} \times \mathcal{B}^* \to (0,1)$ with the feature that for all $w\in \mathcal{B}^*,$ $\tilde{\mu}(1 \, | \, w) + \tilde{\mu}(0\, | \, w)=1.$ In fact, it is more convenient to employ a slightly different representation. Note that any $\tilde{\mu}$ of the above form  induces, for each  $n,$ a probability measure $\mu_n$ on $\mathcal{B}^n.$  Further, for any such $\tilde{\mu}$ and $m\leq n,$ $\mu_m$ and $\mu_n$ are consistent.\footnote{See item (\ref{itemPk}) of Section \ref{subsecMainChar} for the relevant notion of consistency.} So by the Kolmogorov Consistency Theorem, $\tilde{\mu}$  induces a measure $\mu$ on $\mathcal{C}$ (with $\tilde{\mu}$ computable if and only if $\mu$ is).\footnote{For a treatment of the the Kolmogorov Consistency Theorem for the special case of measures on $\mathcal{C},$ see \citep{baez1970c}. For a general treatment, see, e.g., \citep[Chapter V]{parthasarathy2005probability}.} 

\begin{example}
Define $\tilde{\mu}: \mathcal{B} \times \mathcal{B}^* \to (0,1)$  as follows: if $w\in \mathcal{B}^n$ contains $k$ 1's, then $\tilde{\mu}(1 \, | \, w)=\frac{k+1}{n+2}$ and $\tilde{\mu} (0 \, | \, w)=\frac{n-k+1}{n+2}.$ This map satisfies the condition that for all $w\in \mathcal{B}^*,$ $\tilde{\mu}(w,1) + \tilde{\mu}(w,0)=1.$ The corresponding measure is the Laplace--Bayes prior (the Lebesgue--uniform mixture of the Bernoulli measures).
\end{example}
\noindent Not all measures in $\mathcal{P}$ correspond in this way to such $\tilde{\mu}$:  $\mu\in \mathcal{P}$ corresponds to a $\tilde{\mu}$ of the above form if and only if it is a measure of full support (i.e., it assigns positive weight to each open set---or, equivalently, to each basic open set $B_w$). 

\begin{definition}[Forecasters]
A \emph{forecaster} is a Borel probability measure on $\mathcal{C}$ of full support. We denote the family of forecasters by $\mathcal{F}.$     
\end{definition}
\begin{definition}[Forecasting Machines]
A \emph{forecasting machine} is a computable Borel probability measure on $\mathcal{C}$ of full support. We denote the family of forecasting machines by $\mathfrak{F}.$     
\end{definition}

We are going to distinguish three criteria for successful next-chance learning.\footnote{Investigation of inductive learning as next-chance learning traces back to Solomonoff \citep{solomonoff1964formal}. Several criteria of success are prevalent in the literature on Solomonoff induction \citep{solomonoff1978complexity,hutter2007universal,li2013introduction}. But these differ from those considered below in their focus on average or expected performance.\label{fnSolo}} The most restrictive one, due to Blackwell and Dubins \citep{blackwell1962merging}, requires that the forecaster eventually offer answers arbitrarily similar to those of the source concerning any (measurable) question that might be asked about the data stream.\footnote{Note that in the deterministic setting of Section \ref{secLearnSeq} above, the distinction between eventually becoming good at answering all questions and eventually becoming good at predicting the next bit collapsed---recall Proposition \ref{propGuess} above.} The intermediate one, due to Kalai and Lehrer \citep{kalai1994weak}, requires that the forecaster's probabilisitic predictions concerning the next bit eventually approach the true values arbitrarily closely.\footnote{For relations between this criterion of success and those alluded to in fn. \ref{fnSolo} above, see \citep{ryabko2007sequence}.} The least restrictive one, due to Lehrer and Smorodinsky \citep{lehrer1996merging}, relaxes this last requirement by allowing errors, so long as they eventually become arbitrarily rare. 

\begin{definition}[Strong NC-Learning]
We say that forecaster  $\mu$  \emph{strongly NC-learns} source $\lambda$ (or that $\lambda$ is \emph{strongly NC-learnable} by $\mu$) if with $\lambda$-probability 1 the data stream $\sigma \in \mathcal{C}$ satisfies: 
\begin{eqnarray*}
\lim_{n\to \infty} \sup_{A \in \mathfrak{B}} | \mu(A \,|\, \sigma[n]) - \lambda(A  \,|\, \sigma[n])| = 0
\end{eqnarray*}
(recall that $\mathfrak{B}$ denotes the family of Borel subsets of $\mathcal{C}$). \end{definition}

\begin{definition}[NC-Learning]
We say that forecaster  $\mu$ \emph{NC-learns} source $\lambda$ (or that $\lambda$ is \emph{NC-learnable} by $\mu$) if with $\lambda$-probability 1 the data stream $\sigma \in \mathcal{C}$ satisfies 
\begin{eqnarray*}
\lim_{n\to \infty} \mu(s \, | \,\sigma[n]) - \lambda(s \, | \, \sigma[n] ) = 0 \,\,\,\,\,\, s =0,1.
\end{eqnarray*}
\end{definition}

\begin{definition}[Weak NC-Learning]
We say that forecaster  $\mu$ \emph{weakly NC-learns} source $\lambda$ (or that $\lambda$ is \emph{weakly NC-learnable} by $\mu$) if with $\lambda$-probability 1,  the data stream $\sigma \in \mathcal{C}$ satisfies 
\begin{eqnarray*}
\lim_{n\in K \to \infty} \mu(s\, | \, \sigma[n]) - \lambda(s \, | \, \sigma[n] ) = 0 \,\,\,\,\,\, s =0,1
\end{eqnarray*}
for some  $K \subset \mathbb{N}$ with asymptotic density one.
\end{definition}

\begin{remark}[Weaker Variants of Weak NC-Learning.] In parallel with the definition of NV($r$)-learning (see Remark \ref{remPodniekKinber} above), we could introduce, for each $r \in (0,1]$ a notion of NC($r$)-learning, by altering the definition of weak NC-learning to require $K$ to have asymptotic density at least $r.$ All of the propositions below continue to hold if `weak NC-learning' is replaced by `NC($r$)-learning' for any $r\in (0,1]$ (except Proposition \ref{propMeasWeakEvil}, which requires the restriction $r> 1/2$).
\end{remark}

\begin{example}
If $\mu$ is a forecaster, then $\mu$ is also a source and it is immediate that $\mu$ strongly NC-learns, NC-learns, and weakly NC-learns $\mu.$    
\end{example}

\begin{proposition}[Kalai, Lehrer, and Smorodinsky \citep{kalai1994weak,lehrer1996merging}] For any source $\lambda$ and any forecaster $\mu,$ strong NC-learnability of $\lambda$ by $\mu$ implies NC-learnability (but not conversely) and NC-learnability of $\lambda$ by $\mu$ implies weak NC-learnability (but not conversely). 
\proof Strong NC-learnability implies NC-learnability: in the definition of strong NC-learnabil\-ity, for each $n$ take $A$ to be the event of the $(n+1)$st bit being a 1. To see that the converse is not true, consider the family $\{ \lambda_p \,|\, p \in (0,1) \}$ of (non-extreme) Bernoulli measures and let $\mu$ be the Laplace--Bayes prior. It is a basic fact about $\mu$ that it is statistically consistent for the problem of identifying the bias of a coin from knowledge of outcomes of a sequence of tosses \citep{freedman1963asymptotic}. It follows that the forecaster $\mu$ NC-learns each $\lambda_p.$  But $\mu$ does not strongly NC-learn any $\lambda_p$: let $E_p$ be the event that the limiting relative frequency of 1's in the data stream is $p$; then $\lambda_p(E_p)=1$ but $\mu(E_p)=0$; so for any $w\in \mathcal{B}^*,$  $|\mu(E_p\, | \, w) - \lambda_p(E_p\, | \, w)|=1.$  \\
Clearly, NC-learnability implies weak NC-learnability. To see that the converse is not true, take $\mu$ to be the fair coin measure and take $\lambda$ to be the source that generates bits $s_1,$ $s_2,$ \ldots\  as follows: for $k=10^m,$ $s_k$ is the $m$th bit in the binary expansion of $\pi$; all other $s_j$ are generated by flipping a fair coin. The forecaster $\mu$ weakly NC-learns this $\lambda$ but does not NC-learn it, since there are large discrepancies between the forecast probabilities and the true probabilities at arbitrarily late times.    
\hfill \qed
\end{proposition}

We are going to see that relative to each of these three criteria, the problem of next-chance learning  is formidably difficult and involves hard choices.

\subsection{Strong NC-learning} \label{subsecStrongNC}

A famous result and its converse give a necessary and sufficient condition for a source to be strongly NC-learnable by a forecaster.

\begin{proposition}[Blackwell and Dubins \citep{blackwell1962merging}] \label{propBlackwellDubins}
If source $\lambda$ is absolutely continuous with respect to forecaster $\mu$ (i.e., $\lambda(A)>0$ implies $\mu(A)>0$ for all $A\in \mathfrak{B}$), then $\mu$ strongly NC-learns $\lambda.$ 
 \end{proposition}

\begin{proposition}[Lehrer and Smorodinsky \citep{lehrer1996merging}] If forecaster $\mu$ strongly NC-learns source $\lambda,$ then $\lambda$ is absolutely continuous with respect to  $\mu.$ \label{propAbsCont2} 
\end{proposition}

\begin{proposition} Let $\mu$ be a forecaster. The sources strongly NC-learnable by $\mu$ form a dense subset of $\mathcal{P}$ of cardinality $\mathfrak{c}.$ \label{propStrongDense}
\proof Let $w_1,$ \ldots, $w_n$ be binary strings such that $\mathcal{C}$ is a disjoint union of the $B_{w_k}.$ And let $p_1,$ \ldots\ $p_n \in (0,1)$ with $\sum_{k=1}^np_k=1.$ Since the $B_{w_k}$ partition $\mathcal{C},$ each $w \in \mathcal{B}$ is either one of the $w_k,$ or a proper prefix of some of the $w_k,$ or a proper extension of one of the $w_k.$ We define a map $\bar{\lambda}: \mathcal{B}^* \to [0,1]$ as follows:
\begin{enumerate}[(a)]
\item  If $w=w_k$ for some $k,$ then $\bar{\lambda}(w)=p_k.$ 
\item  If $w$ is a prefix of $w_{j_1}, \ldots, w_{j_\ell},$ then $\bar{\lambda} (w)=\sum_{k=1}^\ell p_{j_k}.$
\item  If $w$ is of the form $w_k.v$ for some binary string $v,$ then  $\bar{\lambda}(w)=p_k \cdot \mu(v \, | \, w_k).$
\end{enumerate}
It is immediate that $\bar{\lambda}(\varnothing)=1$ and that $\bar{\lambda}(w)=\bar{\lambda}(w.0)+\bar{\lambda}(w.1)$ for each $w\in \mathcal{B}^*.$ So by the Carath\'eodory Extension Theorem, $\bar{\lambda}$ extends uniquely to a measure $\lambda \in \mathcal{C}$ such that $\lambda(w)=\bar{\lambda}(w)$ for each $w \in \mathcal{B}^*.$ \\
The source $\lambda$ is strongly NC-learnable by $\mu.$ For suppose that $A$ is a Borel subset of $\mathcal{C}$ with $\lambda(A)>0.$ Given the law of total probability, 
\begin{eqnarray*}
\lambda (A) & = & \sum_{k=1}^n \lambda (A\, | \, w_k) \lambda (w_k),
\end{eqnarray*}
there must be some $1\leq \ell \leq n$ such that  $\lambda (A\, | \, w_\ell) > 0.$ It follows that $\mu (A \, | \, w_\ell)>0.$ And since $\mu(w_\ell)$ is also positive ($\mu$ being a forecaster) we find that $\mu(A) >0$ (by the law of total probability, again). So $\lambda$ is absolutely continuous with respect to $\mu$ and Proposition \ref{propBlackwellDubins} tells us that $\mu$ strongly NC-learns $\lambda.$ And since the $w_k$ and the $p_k$ can be chosen arbitrarily, we construct in this way continuum-many such sources in any finite intersection of sub-basic open sets of $\mathcal{P}.$  \hfill \qed
\end{proposition}

The next result follows from the stronger Proposition \ref{propWeakco-meagre} below, but we include it here in order to indicate an independent route to establishing it.

\begin{proposition}[Noguchi \citep{noguchi2015merging}] For any $\mu \in \mathcal{F},$ the set $S_\mu \subset \mathcal{P}$ of sources strongly NC-learned by $\mu$ is meagre in $\mathcal{P}.$ \label{propSNCmeagre}
\proof A classical result tells us that for any measure in $\mathcal{P},$ there is some meagre subset of $\mathcal{C}$ to which it assigns probability 1.\footnote{Szpilrajn \citep{szpilrajn1934remarques} shows that any non-atomic Borel probability measure on a separable metric space assigns measure 0 to some co-meagre set.  Marczewski (=Szpilrajn) and  Sikorski \citep{marczewski1949remarks} observe that this result implies that in a separable metric space without isolated points, every Borel probability measure assigns probability 0 to some co-meagre set. In fact, the hypothesis of separability can be dropped  \citep[Corollary 3.7]{zindulka1999killing}.} And Proposition 1 of \citep{dekel2006non} tells us that for any meagre subset of $\mathcal{C},$ the set of probability measures that assign it positive probability is meagre in $\mathcal{P}.$\footnote{This is a special case of a result of  Koumoullis \citep{koumoullis1996baire}.} So let $A$ be a meagre subset of $\mathcal{C}$ such that $\mu(A)=1$ and let $P_A\subset \mathcal{P}$ be the, necessarily meagre, set of measures that assigns $A$ positive probability. By Proposition \ref{propAbsCont2}, if $\lambda \in S_\mu,$ then $\lambda \in P_A.$ So $S_\mu,$ being a subset of a meagre set, is meagre.
\hfill \qed \end{proposition}
\begin{corollary}
The set $\{ \lambda \in \mathcal{P} \,\, | \,\, \exists \mu \in \mathfrak{F} \mbox{ such that } \mu \mbox{ strongly NC-learns } \lambda\}$ is meagre in $\mathcal{P}.$ 
\end{corollary}

\begin{proposition} For any $\mu \in \mathcal{F},$ the set $J_\mu$ of forecasters that fail to strongly NC-learn any sources strongly NC-learned by $\mu$ is co-meagre in $\mathcal{P}.$ \label{propStrongEvil}
\proof Let $N_\mu$ be the set of $\nu \in \mathcal{P}$ such that there is no $\lambda \in \mathcal{P}$ that is absolutely continuous with respect to both $\mu$ and $\nu.$ By Propositions \ref{propBlackwellDubins} and \ref{propAbsCont2},  $J_\mu \subseteq N_\mu \bigcap \mathcal{F}.$ So it suffices to show that $N_\mu$ and $\mathcal{F}$ are both co-meagre subsets of $\mathcal{P}.$ \\
$N_\mu$ is co-meagre. Let $A$ and $P_A$ be as in the proof of the preceding proposition. Suppose that $\nu \in \mathcal{P}$ is not in $N_\mu.$ So there is a $\lambda \in \mathcal{P}$ absolutely continuous with respect to both $\mu$ and $\nu.$ So $\lambda$ must assign the complement of $A$ zero probability (since $\mu$ does), which means that $\nu$ must assign $A$ positive probability (since $\lambda$ does)---so $\lambda$ is in the complement of the co-meagre set $\mathcal{P}_A.$ So the complement of $N_\mu$ is meagre, being a subset of the meagre set $P_A.$ \\
$\mathcal{F}$ is co-meagre. The forecasters form a dense $G_\delta$ subset of $\mathcal{P}$ \citep[\S 3.13]{dubins1964measurable}.  And in any completely metrizable space (such as $\mathcal{P}$), any dense $G_\delta$ subset is co-meagre \citep[Theorem 9.2]{oxtoby2013measure}.
\hfill \qed \end{proposition}
\begin{corollary}
For any forecaster, there is another, such that the sets of sources strongly NC-learned by the two forecasters are disjoint. 
\end{corollary}

 As usual, we call a sequence $\{ \lambda_i \}_{i \in \mathbb{N}}$ of elements of $\mathfrak{P}$ \emph{uniformly computable in $i$} if there is a computable $F: \mathbb{N} \times \mathcal{B}^* \times \mathbb{N} \to \mathbb{Q}$ such that $| \lambda_i (w) - F(i,w,n) | \leq 2^{-n},$ for all $i,n \in \mathbb{N}$ and $w \in \mathcal{B}^*.$ 

\begin{proposition} (a) Let $\mu \in \mathcal{F}$ and let $S$ be a countable subset of $\mathcal{P}.$ Then there is a  $\mu^* \in \mathcal{F}$ that strongly NC-learns every source in $S$ as well as every source strongly NC-learned by $\mu.$ (b)  Let $\mu \in \mathfrak{F}$ and let $S = \{ \lambda_i \}_{i \in \mathbb{N}}$ be a sequence of measures in $\mathfrak{P}$ that is uniformly computable in $i.$ Then there is a  $\mu^* \in \mathfrak{F}$ that strongly NC-learns every source in $S$ as well as every source strongly NC-learned by $\mu.$ \label{propStrongLearnBetter}
\proof For part (a), enumerate the members of $S$: $\lambda_1,$ $\lambda_2,$ \ldots\  and set 
\begin{eqnarray*}
\mu^* =\frac{1}{2} \mu +  \frac{1}{2} \sum_{k=1}^\infty \frac{1}{2^k} \lambda_k.
\end{eqnarray*}
Each $\lambda_k$ is absolutely continuous with respect to $\mu^*,$ so by Proposition \ref{propBlackwellDubins}, $\mu^*$ strongly NC-learns every source in $S.$ And if $\nu$ is a source strongly NC-learned by $\mu,$ then by Proposition \ref{propAbsCont2}, $\nu$ must be absolutely continuous with respect to $\mu$ and hence also with respect to $\mu^*$---so by Proposition \ref{propBlackwellDubins}, $\mu^*$  strongly NC-learns $\nu.$ \\
For part (b), we can proceed in the same way. The only thing to check is that  if $\mu \in \mathfrak{P}$ is computable and $\{ \lambda_i \} \subset \mathfrak{P}$ is uniformly computable in $i,$ then the measure $\mu^*$ as defined above is also computable.  To this end, suppose that $F_0: \mathcal{B}^* \times \mathbb{N} \to \mathbb{Q}$ and $F: \mathbb{N} \times \mathcal{B}^* \times \mathbb{N} \to \mathbb{Q}$ are computable, with $|\mu(w) - F_0(w,n)| \leq 2^{-n}$ and $|\lambda_i(w) - F(i,w,n)| \leq 2^{-n},$ for all $w \in \mathcal{B}^*$ and $i,n\in \mathbb{N}.$ \\
We define $F^* : \mathcal{B}^* \times \mathbb{N} \to \mathbb{Q}$ as follows:  
\begin{eqnarray*}
F^*(n,w) & := & \frac{1}{2} F_0(w,n+1) + \frac{1}{2} \sum_{k=1}^{n+1} \frac{1}{2^{k}}F(k, w, 2n) 
\end{eqnarray*}
Then for any given $w \in \mathcal{B}^*$ and $n \in \mathbb{N},$ we define $\alpha, \beta, \gamma \in \mathbb{R}$:
\begin{eqnarray*}
\alpha & := &  \frac{1}{2} \left( \mu(w) - F_0(w,n+1) \right) \\
\beta & := & \sum_{k=1}^{n+1} \frac{1}{2^{k+1}}\left(\lambda_k(w) - F(k, w, 2n)\right) \\
\gamma & := & \sum_{k=n+2}^{\infty} \frac{1}{2^{k+1}} \lambda_k(w).
\end{eqnarray*}
Note that each of $|\alpha|,$ $|\beta|,$ and $|\gamma|$ is no greater than $2^{-(n+2)}.$ In the case of $|\alpha|,$ this follows from what we know about $F_0.$ For $|\beta|,$ we have:
\begin{eqnarray*}
| \beta | & \leq & \sum_{k=1}^{n+1} \frac{1}{2^{k+1}}\left| \lambda_k(w) - F(k, w, 2n)\right| \\
& \leq & \sum_{k=1}^{n+1} \frac{1}{4}\left| \lambda_k(w) - F(k, w, 2n)\right| \\
& \leq & \sum_{k=1}^{n+1} \frac{1}{4} 2^{-2n} \\
& = & \frac{n+1}{2^{n}}2^{-(n+2)}, 
\end{eqnarray*}
and for any $n\geq 1$ we have that $n+1 \leq 2^n.$ And since for each $k$ we have $0 \leq \lambda_k(w) \leq 1,$ we have that $| \gamma |  \leq  \sum_{k=n+2}^{\infty} \frac{1}{2^{k+1}}.$ \\
Now, $F^*$ is computable and we have:
\begin{eqnarray*} 
|\mu^*(w) - F^*(w,n)| & = &  \left| \alpha + \beta + \gamma \right| \\
& \leq &  |\alpha| + |\beta| + |\gamma|  \\
& < & 2^{-n}. 
\end{eqnarray*} 
So $\mu^*$ is computable. \hfill \qed
\end{proposition}

Thus we have a no-free-lunch result for strong NC-learning: the set of sources strongly NC-learned by any forecaster is uncountable and dense but meagre; for every (computable) forecaster there is another (computable) forecaster that strongly NC-learns everything it does, plus a further countably infinite set of sources; and for every forecaster there is another that strongly NC-learns a disjoint set of sources (indeed, typical forecasters have this feature).

\subsection{NC-Learning and Weak NC-Learning} \label{subsecWeakMeas}

Proposition \ref{propSNCmeagre} above tells us that each forecaster strongly NC-learns a dense and uncountable but meagre set of sources. This implies that the sets of sources NC-learned and weakly NC-learned by any forecaster are also dense and uncountable. Strong NC-learning is, intuitively, a much more restrictive notion than NC-learning: being able to accurately answer all questions about the data stream, including questions about the infinite future, is much a much more demanding standard than being able to accurately estimate the chances for the next bit.\footnote{Indeed, there is a sense in which strong NC-learning implies \emph{rapid} NC-learning, and a sense in which the converse implication holds  \citep[Propositions 2 and 3 ]{sandroni1999speed}.\label{fnSandroni}} Similarly, NC-learning is, intuitively, a much more restrictive notion than weak NC-learning: we saw in Section \ref{secLearnSeq} above that weakening NV-learning by allowing an infinite number of errors (so long as they were of asymptotic density zero) made a marked difference to the size of the set of sequences that a given extrapolator could learn---any extrapolator NV-learns a countable set of sequences but weakly NV-learns an uncountable set of sequences. So it is not obvious that the set of sources (weakly) NC-learned by a given forecaster should always be meagre.\footnote{Noguchi \citep[p. 433]{noguchi2015merging}, after discussing the results cited above in fn. \ref{fnSandroni}, remarks that: ``These results lead us to conjecture that, in general, a merged set (of probability measures) [i.e., a set of sources strongly NC-learned by a given forecaster] may be much smaller than a weakly merged set [i.e, a set of sources NC-learned by a given forecaster].'' He then goes on to observe that each forecaster strongly NC-learns a meagre set of sources---so it is natural to read him as conjecturing that the set of sources NC-learned by a forecaster need not be meagre.} Not obvious---but, as we will see, nonetheless true.  

We begin by introducing a basis $\mathcal{W}$ for the weak topology on $\mathcal{P}.$ First, for each $k\in \mathbb{N},$ we fix a metric on $\mathcal{P}_k$ compatible with its topology: we take  the distance between $\lambda, \mu \in \mathcal{P}_k$ to be given by: 
\begin{eqnarray*}
d(\mu,\lambda) := \max_{w \in \mathcal{B}^k} | \mu(w) - \lambda(w) |.
\end{eqnarray*}
In terms of our identification of $\mathcal{P}_k$ with a closed subset of $\mathbb{R}^{2^k},$ this is the metric induced by the $\ell^\infty$ norm on $\mathbb{R}^{2^k}.$ For $\mu \in \mathcal{P}_k$ and $\varepsilon>0,$ we write $B(\mu,\varepsilon)$ for the open metric ball of radius $\varepsilon$ centred at $\mu$:
\begin{eqnarray*}
B(\mu,\varepsilon):= \{ \lambda \in \mathcal{P}_k \, | \, d(\mu,\lambda)<\varepsilon \}.
\end{eqnarray*}
We call $B(\mu,\varepsilon)\subset  \mathcal{P}_k$ \emph{rational} if $\varepsilon \in \mathbb{Q}$ and $\mu(w) \in \mathbb{Q}$ for each $w\in \mathcal{B}^k.$ 

For each $k \in \mathbb{N},$ let $\Pi_k: \mathcal{P} \to \mathcal{P}_k$ be the restriction map: for $\mu \in \mathcal{P},$ $\Pi_k(\mu)$ is the measure in $\mathcal{P}_k$ such that $\Pi_k(\mu)(w)=\mu(w)$ for each $w \in \mathcal{B}^k.$ We now take $\mathcal{W}$ to comprise the inverse images under the $\Pi_k$ of the rational open metric balls in the various $\mathcal{P}_k$: 
\begin{eqnarray*}
\mathcal{W} := \{ W=\Pi_k^{-1}(B(\mu, \varepsilon)) \, | \, k\in \mathbb{N},\,  \mu \in \mathcal{P}_k, \, \mu(w)\in \mathbb{Q} \,\, \forall w\in \mathcal{B}^k, \, \varepsilon>0, \, \varepsilon \in \mathbb{Q} \}.
\end{eqnarray*}

\begin{proposition}
$\mathcal{W}$ is a basis for the weak topology on $\mathcal{P}.$
\proof It suffices to show: (i) that each $W \in \mathcal{W}$ is open; and (ii) that for any non-empty open set $U \subset \mathcal{P}$ and for any $\nu \in U,$ there is a $W \in \mathcal{W}$ with $\nu \in W \subset U.$ 

\noindent (i) Fix $W\in \mathcal{W}$ of the form $W = \Pi_k^{-1}(B(\mu, \varepsilon)).$ Let $w_1,$ \ldots, $w_{2^k}$ be an enumeration of the $k$-bit strings. And for each $1\leq j \leq 2^k,$ let $p_j:= \max \{0, \mu(w_j) - \varepsilon\}$ and $q_j:= \min \{1, \mu(w_j) + \varepsilon\}.$ Then we have:
\begin{eqnarray*}
W & = & \Big\{ \lambda \in \mathcal{P} \, | \, \max_{1\leq j \leq 2^k} | \mu(w_j) - \lambda(w_j) | < \varepsilon  \Big\} \\
& = & \bigcap_{j=1}^{2^k} S_{w_j,p_j,q_j},
\end{eqnarray*}
where each $S_{w_j,p_j,q_j}$ is a sub-basic open subset of $\mathcal{P}$ (as in item (\ref{itemSubBasic}) of Section \ref{subsecMainChar} above). So $W$ is an open subset of $\mathcal{P}.$

\noindent (ii) It suffices to consider a non-empty open set $U\subset \mathcal{P}$ that is a finite intersection of sub-basic open sets. Let $S_{w_1,p_1,q_1}, \ldots , S_{w_n,p_n,q_n}$ be arbitrary sub-basic open subsets of $\mathcal{P}$ and suppose that $U:=\bigcap_{k=1}^nS_{w_k,p_k,q_k}\neq \varnothing.$ Let $N = \max \{|w_1|, \ldots, |w_n|\}$ and let $\nu \in U.$ Note that each $\Pi_N(S_{w_j,p_j,q_j})$ is an open subset of $\mathcal{P}_N$: each condition of the form $p_j < \Pi_N(\mu)(w_j) < q_j$  just imposes an inequality on (sums of) differences of coordinate relative to our identification of $\mathcal{P}_N$ with a subset of $\mathbb{R}^{2^N}.$ So we can find a rational open metric ball $B$ contained in $\Pi_N(U)$ with $\Pi_n(\nu)\in B.$ Letting $W:=\Pi_N^{-1}(B) \in \mathcal{W},$ we have $\nu \in W \subset U.$  
\hfill \qed
\end{proposition}

\begin{proposition} \label{propWeakco-meagre}
Let $\mu$ be a forecaster. The sources weakly NC-learnable by $\mu$ form a meagre subset of $\mathcal{P}.$ 
\proof For any source $\lambda$ and $k \in \mathbb{N},$ let us say that $(\mu,\lambda)$ considers $k$ \emph{bad} if for each $w \in \mathcal{B}^k$ we have 
\begin{eqnarray*}
|\mu(s\, | \, w) - \lambda(s\, | \, w)| \geq \frac{1}{5} \hspace{1cm} s=0,1.
\end{eqnarray*}
And let us say that $(\mu,\lambda)$ considers $k$ \emph{super-bad} if $(\mu,\lambda)$ considers more than half of the $j \leq k$ to be bad. And, by extension, for any subset $S \subset \mathcal{P},$ let us say that $(\mu,S)$ considers $k \in \mathbb{N}$ (super-)bad if $(\mu,\lambda)$ does for each $\lambda \in S.$ 

For each $n \in \mathbb{N},$ let $F_n$ be the set of $\lambda \in \mathcal{P}$ such that $(\mu, \lambda)$ considers at least $n$ natural numbers to be super-bad. And let $A_n$ be the complement of $F_n$ in $\mathcal{P}.$ Note that $\mu$ cannot NC-learn $\lambda$ if there are infinitely many $k \in \mathbb{N}$ that $(\mu,\lambda)$ considers bad and that $\mu$ cannot weakly NC-learn $\lambda$ if there are infinitely many $k \in \mathbb{N}$ that $(\mu,\lambda)$ considers super-bad. So if $\mu$ weakly NC-learns $\lambda,$ then $(\mu, \lambda)$ can consider only finitely many natural numbers to be super-bad, which means that there will be an $N$ such that $\lambda \notin F_N,$ which implies that $\lambda \in A:= \bigcup_{n\in \mathbb{N}} A_n.$ So in order to establish our proposition, it suffices to show that each $A_n$ is nowhere dense in $\mathcal{P}.$

The first step is to suppose that we are given a set $W \in \mathcal{W} $ of the form $\Pi_k^{-1}(B(\lambda, \varepsilon))$ and to show how to find $W_1 \in \mathcal{W}$ of the form $W_1=\Pi_{k+1}^{-1}(B(\lambda_1, \varepsilon_1)),$ such that  $W_1 \subset W$ and $(\mu,W_1)$ considers $k+1$ to be bad. 

First, we select $\lambda_1.$ For each $w \in \mathcal{B}^k,$ if $\mu (w.0)\geq \mu (w.1),$ we set  
\begin{eqnarray*}
\lambda_{1}(w.0)= \frac{1}{10} \cdot \lambda (w) \hspace{1cm} \mbox{and} \hspace{1cm}  \lambda_{1}(w.1)= \frac{9}{10} \cdot \lambda (w);
\end{eqnarray*}
otherwise we set  
\begin{eqnarray*}
\lambda_{1}(w.0)= \frac{9}{10} \cdot \lambda (w) \hspace{1cm} \mbox{and} \hspace{1cm} \lambda_{1}(w.1)= \frac{1}{10} \cdot \lambda (w).
\end{eqnarray*}
This gives us a well-defined $\lambda_{1} \in \mathcal{P}_{k+1}$ that assigns rational values to each string in $\mathcal{B}^{k+1}.$  

We now select $\varepsilon_{1}>0$ as follows: we choose $m$ large enough so that $\varepsilon_{1}=10^{-m}$ is small enough so that for any $\lambda'$ in $B(\lambda_1,\varepsilon_1),$ for each $w \in \mathcal{B}^k,$ if $\mu (w.0)\geq \mu (w.1),$ then  
\begin{eqnarray*}
  \lambda'(w.1) > \frac{8}{10} \cdot \lambda (w),
\end{eqnarray*}
and if  $\mu (w.0)< \mu (w.1)$ then,
\begin{eqnarray*}
\lambda'(w.0) >  \frac{8}{10}\cdot \lambda (w).
\end{eqnarray*}
This process can be iterated. In particular, if we are given $W \in \mathcal{W}$ of the form  $\Pi_k^{-1}(B(\lambda, \varepsilon)),$ we can run the process once to construct $W_1 \in \mathcal{W}$ with $W_1 \subset W$ such that $(\mu, W_1)$ considers $k+1$ bad; applying the process again (now with $W_1$ if place of $W$) yields a $W_2\in \mathcal{W}$ with $W_2\subset W_1$ such that $(\mu, W_2)$ considers $k+2$ bad and so on.

So if we are given $W \in \mathcal{W}$ of the form $\Pi_k^{-1}(B(\lambda, \varepsilon)),$ we can run the process $k+n$ times to yield $W^*:=W_{k+n} \in \mathcal{W}$ such that $W^*\subset W$ and $(\mu, W^*)$ considers at least $n$ numbers to be super-bad, so that $W^* \bigcap A_n= \varnothing.$ So $A_n$ is nowhere dense in $\mathcal{P}.$ 
\hfill \qed
\end{proposition}
\begin{corollary} The set $\{ \lambda \in \mathcal{P} \,\, |\,\, \exists \mu \in \mathfrak{F} \mbox{ such that } \mu \mbox{ weakly NC-learns }\lambda \}$ is meagre in $\mathcal{P}.$  
\end{corollary}

It of course follows that the set of sources (strongly) NC-learnable by a given forecaster are likewise meagre---and that set of all sources collectively (strongly) learnable by forecasting machines are likewise meagre. 

\begin{remark} The set of sources not even weakly NC-learnable by a given forecaster  $\mu$ is a co-meagre subset of $\mathcal{P},$ and so is uncountable. A variant on the proof of the above proposition shows that even if the continuum hypothesis fails, this set has the cardinality of the continuum.  Fix a metric on $\mathcal{P}$ compatible with the weak topology, such as the Prokhorov metric.\footnote{For details see, e.g., \citep[pp. 72 f.]{Billingsley:1999aa}.} And let us amend the iterative procedure of the proof of the preceding proposition so that the diameter of $W_{k+1}$ relative to this metric is no more than half of the diameter of $W_k.$ Then if we are given $W \in \mathcal{W}$ and repeatedly apply our revised iterative procedure, we will construct a sequence $W_1,$ $W_2,$ \ldots\ of nested $\mathcal{W}$-sets such that $\bigcap_{k=1}^\infty W_k$ contains a single source, which is not even weakly NC-learnable by $\mu.$ There are continuum-many distinct $W$ we could use to initiate this procedure---these determine continuum-many distinct sources not even weakly NC-learnable by $\mu.$
\end{remark}

\begin{proposition} Let $\mu$ be a forecaster. Then there is a second forecaster $\mu^\dagger$ such that the sets of sources weakly NC-learned by $\mu$ and by $\mu^\dagger$ are disjoint. If $\mu$ is computable, we can take $\mu^\dagger$ to be likewise computable. \label{propMeasWeakEvil} 
\proof Let $\mu$ be given. We construct a map $\nu:  \mathcal{B}^* \to [0,1]$ inductively as follows:
\begin{enumerate}[a)]
\item $\nu(\varnothing)=1$;
\item Supposing that $\nu(w)$ is given, we define $\nu(w.0)$ and $\nu(w.1)$ as follows:
\begin{enumerate}[i)]
\item if $\mu(0\, |\, w)\leq \mu(1\, | \, w)$, then $\nu(w.0)=9/10 \cdot \nu(w)$ and $\nu(w.1)=1/10 \cdot \nu(w)$;
\item if $\mu(0\, |\, w)> \mu(1\, | \, w)$, then $\nu(w.0)=1/10 \cdot \nu(w)$ and $\nu(w.1)=9/10 \cdot \nu(w).$
\end{enumerate}
\end{enumerate}
Clearly, for any $w\in \mathcal{B}^*,$ $\nu(w)= \nu(w.0)+\nu(w.1).$ So by the Carath\'eodory Extension Theorem, $\nu$ extends to a unique Borel probability measure on $\mathcal{C},$ which we take as our $\mu^\dagger.$  \\
For any non-empty $w \in \mathcal{B}^*,$ $| \mu(1 \, |\, w) - \mu^\dagger(1 \, |\, w)| \geq \frac{2}{5}.$ So for any $\lambda \in \mathcal{P},$ any $\sigma \in \mathcal{C},$ and $n\in \mathbb{N}$ we have:
\begin{eqnarray*}
 \max \{ |\mu(1\, |\, \sigma[n]) - \lambda (1 \, | \, \sigma[n])|, \,\, |\mu^\dagger(1\, | \,\sigma[n]) - \lambda (1 \, | \, \sigma[n])| \}
& \geq & \frac{1}{5}.
\end{eqnarray*}
So there can be no $\lambda\in \mathcal{P}$ such that for every $\sigma$ in a set of $\lambda$-measure one, there is a set $K$ of natural numbers of asymptotic density 1, such that for sufficiently large $n \in K,$ $\mu(1\, | \, \sigma[n])$ and $\mu^\dagger(1\, |\, \sigma[n])$ are both arbitrarily close to $\lambda(1\, |\, \sigma[n])$---i.e.,  there is no source $\lambda$ that is weakly NC-learned by both $\mu$ and $\mu^\dagger.$
\hfill \qed
\end{proposition}
 
 \noindent Of course, it follows that $\mu$ and $\mu^\dagger$ also (strongly) NC-learn disjoint sets of sources. 
  
 So we have no-free-lunch theorems for (weak) NC-learning: each forecaster, computable or not, (weakly) NC-learns an uncountable and dense but meagre set of sources; and for each (computable) forecaster there is another that (weakly) NC-learns a disjoint set of sources. 
 
\begin{remark}
Lehrer and Smorodinsky \citep{lehrer1996merging} show that if $\mu \in \mathcal{F}$ NC-learns $\lambda \in \mathcal{P},$ then any nontrivial mixture of $\mu$ with any $\nu \in \mathcal{P}$ weakly NC-learns $\lambda.$ Ryabko and Hutter \citep[Proposition 10]{ryabko2007sequence} show that this result is sharp: they given an example of of measures $\mu,$ $\nu,$ and $\lambda$ where $\mu$ NC-learns $\lambda$ but any non-trivial mixture of $\mu$ and $\nu$ merely weakly NC-learns $\lambda.$ So there is no prospect of using the strategy of the proof of Proposition \ref{propStrongLearnBetter} above to prove an analogous result for NC-learning. 
\end{remark}

\subsection{Forecasting of Computable Sources} \label{secComp} 

Let us now specialize to problem of (strong, weak) NC-learning for computable forecasters facing data streams generated by computable sources.\footnote{Vit\'anyi and Chater  \citep{vitanyi2017identification} introduce a model of learning in which agents facing a data stream generated by a computable source attempt to guess a code number for that source (so this model of learning stands to next chance learning as identification stands to extrapolation). See also \citep{bienvenu2018algorithmic,barmpalias2018equivalences}.}

For $\mu \in \mathfrak{F},$  we denote by $\mathcal{NC}(\mu)$ the set of computable sources that are NC-learned by $\mu.$  We use $\mathcal{NC}$ to denote:
\begin{eqnarray*}
\{ S \subset \mathfrak{P} \,\, | \,\, \exists \mu \in \mathfrak{F} \mbox{ with } S \subseteq NC(\mu) \}.
\end{eqnarray*}
Let us likewise use $\mathcal{NC}^s(\mu)$ and $\mathcal{NC}^w(\mu)$ to denote the set of computable sources strongly NC-learned  and weakly NC-learned by the forecasting machine $\mu$ and use $\mathcal{NC}^s$ and $\mathcal{NC}^w$ to denote the class of subsets of $\mathfrak{P}$ that can be strongly/weakly NC-learned by some forecasting machine. It is immediate from the definitions that $\mathcal{NC}^s\subseteq\mathcal{NC}$ and that $\mathcal{NC}\subseteq\mathcal{NC}^w.$ A variant on the proof of Proposition \ref{exBC} shows that the latter containment is proper. 

\begin{proposition} 
$\mathcal{NC}\subset\mathcal{NC}^w.$
\proof Let $V$ be the subset of $\mathfrak{P}$ consisting of $\delta$-function measures concentrated on computable binary sequences in which 0's have vanishing asymptotic density. Let $\mu \in \mathfrak{F}$ be the measure that on input of any $w\in \mathcal{B}^n,$ considers the chance of seeing a 0 next to be $2^{-n}.$ We have $V \in \mathcal{NC}^w(\mu).$ But suppose that $V \subseteq \mathcal{NC}(\nu)$ for some $\nu \in \mathcal{F}.$ We define $\sigma \in \mathfrak{C}$ as follows: $\sigma$ is of the form $1^{n_1}.0.1^{n_2}.0.1^{n_3}.0\ldots$ where where each $n_j$ is chosen to be the smallest $n$ larger than $2^j$ such that $\nu(1^{n_1}.0.1^{n_2}.0.\ldots. 1^{n_{j-1}}.01^n)>.9$ (such $n_j$ must exist, since by assumption $\nu$ $\mathcal{NC}$-learns each delta-function measure concentrated on a sequence containing only finitely many 0's). The delta-function measure concentrated on $\sigma$ is in $V.$ But $\sigma \notin \mathcal{NC}(\nu),$ since when fed $\sigma,$ there are infinitely many occasions on which $\nu$ issues forecast probabilities for seeing a 0 next of less than .1, when the true chance is 1. 
\hfill \qed
\end{proposition}

\begin{proposition}
For any $\mu \in \mathfrak{F},$ the following are dense subsets of $\mathcal{P}$: 
\begin{enumerate}[(a)]
\item $\mathcal{NC}^s(\mu).$
\item The complement of $\mathcal{NC}^s(\mu)$ in $\mathfrak{P}.$
\item $\mathcal{NC}(\mu).$
\item The complement of $\mathcal{NC}(\mu)$ in $\mathfrak{P}.$
\item $\mathcal{NC}^w(\mu).$
\item The complement of $\mathcal{NC}^w(\mu)$ in $\mathfrak{P}.$
\end{enumerate}
\proof The claim concerning (a) follows via straightforward adaptation of the proof of Proposition \ref{propStrongDense}, while that of (f) follows from Proposition \ref{propEffectMeagreMeas} below. The other sets listed are supersets of (a) or (f).
\hfill \qed
\end{proposition}
\noindent For any of our senses of probabilistic learning, for any computable forecaster, that forecaster succeeds in learning a countable infinity of computable sources and fails to learn a countable infinity of computable sources. So we have parity between the learnable and the unlearnable at the level of cardinality. And this parity persists at the level of topology: a version of Sierpi\'nski's Theorem tells us that, up to homeomorphism, there is only one countable  metrizable topological space without isolated points \citep[\S 2]{neumann1985automorphisms}.

But it is straightforward to extend Mehlhorn's notion of an effectively meagre set to the context of $\mathfrak{P},$ with the elements of the basis $\mathcal{W}$ for $\mathcal{P}$ of Section \ref{subsecWeakMeas} playing the role that the basic open sets $B_w$ played in our discussion of effectively meagre subsets of $\mathfrak{C}$ in Section \ref{subsecCompSeq} above. And, as in the case of next-value learning, we find that the for our species of next-chance learning, this notion allows us to isolate a sense in which failure is incomparably more common than success. 

Recall that elements of $\mathcal{W}$ are specified by specifying a natural number $k,$  a rational-valued measure $\mu \in \mathcal{P}_k$ (which is determined in turn by specifying the values that it assigns each $w \in \mathcal{B}^k$), and a rational $\varepsilon >0.$ So the elements of $\mathcal{W}$ can be effectively represented by binary strings. In the following definition we take such a coding scheme to be fixed.  

\begin{definition}
Let $A$ be a subset of $\mathfrak{P}.$ Let $f: \mathcal{W} \to \mathcal{W}$ be a computable function. Then $A$ is \emph{effectively nowhere dense via $f$} if for each $W\in \mathcal{W}$:
\begin{enumerate}[i)]
\item $f(W) \subset W$;
\item $A \bigcap f(W) =\varnothing.$ 
\end{enumerate}
\end{definition} 

\begin{definition}
A subset $A$ of $\mathfrak{P}$  \emph{effectively nowhere dense} if there is a computable $F: \mathbb{N} \times \mathcal{W} \to \mathcal{W}$ such that:
\begin{enumerate}[i)]
\item for each $n \in \mathbb{N}$ there is an $A_n$ that is effectively nowhere dense via $F(n,\cdot): \mathcal{W} \to \mathcal{W}$;
\item  $A = \bigcup A_n.$
\end{enumerate}
The complement of an effectively meagre subset of $\mathfrak{P}$ is \emph{effectively co-meagre.}
\end{definition}

\noindent The proofs of Propositions \ref{propMehl1} and \ref{propMehl2} are easily adapted to yield:   

\begin{proposition}
The set of effectively meagre subsets of $\mathfrak{P}$ is closed under the following operations:
\begin{enumerate}[(i)]
\item taking subsets;
\item taking finite unions;
\item taking effective unions.
\end{enumerate}
\end{proposition}

\begin{proposition} For any $W \in \mathcal{W},$ the set $W \bigcap \mathfrak{P}$ is not effectively meagre. 
\end{proposition}

\noindent So  it is again natural to consider the elements of effectively meagre subsets of $\mathfrak{P}$ to be incomparably less common than the elements of effectively co-meagre subsets of $\mathfrak{P}.$

\begin{proposition}
For any $\mu \in \mathfrak{F},$ $NC^w(\mu)$ is an effectively meagre subset of $\mathfrak{P}.$ \label{propEffectMeagreMeas}
\proof A straightforward adaptation of the proof of Proposition \ref{propWeakco-meagre}, appealing to the fact that  when $\mu$ is computable, the map $(n,W) \mapsto W^*\notin A_n$ used there is computable.
\hfill \qed
\end{proposition}

\noindent So there is a natural sense in which, for any computable forecaster $\mu,$ among computable sources, those weakly NC-learnable by $\mu$ are incomparably less common than those not NC-weakly learnable by $\mu.$ And, \emph{a fortiori,} those computable sources (strongly) NC-learnable by $\mu$ are incomparably less common than those not (strongly) NC-learnable by $\mu.$ Learning in this setting is formidably difficult. And hard choices must be made: the proof of Proposition \ref{propMeasWeakEvil} above shows that each $\mu \in \mathfrak{F}$ has an evil twin $\mu^\dagger \in \mathfrak{F}$ such that the two weakly NC-learn disjoint sets of measures.

\begin{remark}[Partial Forecasting Machines.]
In Remark \ref{remFortnow} above, we saw above that liberalizing our notion of NV-learning of computable sequences by allowing merely partial computable extrapolating machines made an interesting difference: while every set in $\mathcal{NV}$ is effectively meagre, this is not true of every set in  $\mathcal{NV}^{\prime}$ (let alone $\mathcal{NV}\hspace{1pt}^{\prime\prime}$). What is the `partial' analog of NC-learning? 

Recall that a \emph{semi-measure} on $\mathcal{C}$  is a map $\mu : \mathcal{B}^* \to [0,1]$ satisfying  $\mu(\varnothing )\leq 1$ and $\mu(w)\geq \mu(w.0)+\mu(w.1)$ for all $w\in \mathcal{B}^*$ (with equality of course being required for measures). Such a $\mu$ is \emph{lower semi-computable} if there exists a partial computable  $\phi : \mathcal{B}^*\times \mathbb{N} \to [0,1]$ with $\mu(w)=\lim_{\ell\to \infty} \phi(w,\ell)$ and  $\phi (w, k) \leq \phi (w, k+1),$ for all $w\in \mathcal{B}^*$ and $k\in \mathbb{N}.$ According to the approach deriving from \citep{solomonoff1964formal} and \citep{levin1977invariant}, as computable functions stand to partial computable functions, computable measures stand to lower semi-computable semi-measures.\footnote{For full details and motivation see  \citep{li2013introduction}.} This approach leads to a stunning result: there exists a lower semi-computable semi-measure $\mu_S$ that next-chance learns (in various natural senses) every computable measure (there are many such learners, in fact). But this feat is less impressive than it might at first appear \citep{leike2015computability,sterkenburg2017putnam}: while $\mu_S$ is lower semi-computable, the corresponding conditional probability function $\mu_S(1\, | w)$---i.e., the thing we need to calculate in order to make predictions---is merely limit computable (the problem being that the quotient of two lower semi-computable numbers need not be lower semi-computable). 

Here is an alternative approach.\footnote{In what follows, I am indebted to discussions with Tom Sterkenburg.} A \emph{confirmation function} is a map $\mu : \mathcal{B} \times \mathcal{B}^* \to (0,1)$ such that $\mu (1 \, | \, w) + \mu (0 \, | \, w)=1$ (as noted above, the Kolmogorov Consistency Theorem tells us that specifying a confirmation function is the same thing as specifying a forecaster).  A \emph{semi-confirmation function} is  a map $\mu : \mathcal{B} \times \mathcal{B}^* \to (0,1)$ such that $\mu (1 \, | \, w) + \mu (0 \, | \, w)\leq 1.$ We say that a semi-confirmation function $\mu$ is \emph{lower semi-computable} if there is a  partial computable  $\phi : \mathcal{B} \times \mathcal{B}^*\times \mathbb{N} \to [0,1]$ such that or each $s \in \mathcal{B},$  $w\in \mathcal{B}^*,$ and $k\in \mathbb{N}$ we have: (a)  $\mu(s \, | \, w)=\lim_{\ell\to \infty} \phi(s,w,\ell)$; and (b) $\phi (s,w, k) \leq \phi (s,w, k+1).$ We call a lower-semi computable semi-confirmation function a \emph{partial forecasting machine.} We say that the partial forecasting machine $\mu$ \emph{NC\hspace{1pt}$^\prime$-learns} the computable source $\lambda$ if with $\lambda$-probability one, when $\mu$ is fed a data stream $\sigma$ generated by $\lambda,$ we have: (i)  $\mu(1 \, | \sigma[k]) + \mu(0 \, | \, \sigma[k]) = 1$ for each $k$; and $\mu(1 \, | \sigma[k])$ converges to $\lambda(1 \, | \sigma[k])$ as $k \to \infty.$ We say that $\mu$ \emph{NC\hspace{1pt}$\hspace{1pt}^{\prime\prime}$-learns} $\lambda$ if the previous definition holds with clause (i) weakened to allow finitely many exceptions. 

We define the classes $\mathcal{NC}^\prime$ and $\mathcal{NC}\hspace{1pt}^{\prime\prime}$ in the obvious way. 
It is straightforward to show that if a partial extrapolating machine $m$ NV-/NV$^\prime$-/NV$\hspace{1pt}^{\prime\prime}$-learns each sequence in a set $S_0$ of computable sequences, then there is a partial forecasting machine $\mu$ that NC-/NC$^\prime$-/NC$\hspace{1pt}^{\prime\prime}$-learns each delta-function measure corresponding to an element of $S_0$: for $w \in \mathcal{B}^n,$ and $k \in \mathbb{N},$ set $\phi(1,w,k)=\phi(0,w,k)=0$ unless simulating the computation of $m$ on $w$ for $k$ steps shows that $m(w)=s,$ in which case set $\phi(s,w,k)=1-2^{-n}$ and set $\phi(1-s,w,k)=2^{-n}.$
Further, if a partial forecasting machine $\mu$ NC-/NC$^\prime$-/NC$\hspace{1pt}^{\prime\prime}$-learns each member of a set $S_1$ of computable delta-function measures, then there is a partial forecasting machine $m$ that NV-/NV$^\prime$-/NV$\hspace{1pt}^{\prime\prime}$-learns each sequence that is the support of one of the measures in $S_1$: for $w \in \mathcal{B}^n,$ compute the conditional probabilities $\mu(s\,| \, w)$ until one of them is at least a half and take the corresponding bit to be $m(w).$ 
So from the fact that $\mathcal{NV} \subset \mathcal{NV}^\prime \subset \mathcal{NV}\hspace{1pt}^{\prime\prime}$ it follows that  $\mathcal{NC} \subset \mathcal{NC}^\prime \subset \mathcal{NC}\hspace{1pt}^{\prime\prime}.$ Since there are sets in $\mathcal{NV}^\prime$ that are not effectively meagre in $\mathfrak{C},$ it is natural to expect that there are sets in $\mathcal{NC}^\prime$ that are not effectively meagre in $\mathfrak{P}.$ It is known, however, that $\mathfrak{P}\notin \mathcal{NC}\hspace{1pt}^{\prime\prime}$ \citep{sterkenburg2017putnam}.
\end{remark}

\section{Discussion} \label{secDiscuss}

Over the course of the last century, it became widely accepted that successful inductive learning is possible only against a background of biases that favour some hypotheses over others (see, e.g.,   \citep{jeffreys1933probability,kuhn1963structure,chomsky1965aspects,hempel1966philosophy}).
\footnote{Arguably, this theme can be found already in Leibniz---see Item 6 in \citep{leibniz2012philosophical}.} No-free-lunch results substantiate this insight. If we don't presuppose anything about the binary sequence being revealed to us, then we face a formidably difficult learning problem: no matter what approach to learning we adopt, the situations that we might face,  those in which we fail are incomparably more common than those in which we succeed. And no approach dominates all rivals in its range of success: for any approach, there are others that succeed in situations in which the given one fails; indeed, for any approach, there is another that succeeds in a disjoint set of situations.  To adopt an approach to learning is to make a bet about what the world is like.

No-free-lunch results place upper bounds on our reasonable ambitions.\footnote{That is one of their uses. Of the results canvassed in Section \ref{secIntro} above, Putnam's was designed to expose a serious flaw in objective Bayesian approaches in the tradition of \citep{carnap1945inductive} while that of Shalev--Shwartz and Ben--David was devised to provide an elegant motivation for the definition of VC-dimension.} Suppose that one is interested in the question: Why should someone interested in arriving at the truth proceed inductively (expecting the future to be like the past) rather than counter-inductively? Consider how this question looks in the simplest of our contexts, in which an agent being shown a binary sequence bit by bit aims to eventually be able to correctly predict each new bit on the basis of the bits seen so far. Here each method of learning succeeds on a countable dense subspace of the space of binary sequences. And all such subspaces are isomorphic (Sierpi\'nski's Theorem again). So unless we impose more structure on our problem, we have parity between the set of possibilities in which a inductive extrapolator $m$ is successful and the set of possibilities in which counter-inductive extrapolator $m^\dagger$ is successful.\footnote{Some formal learning theorists take the view that learning strategies prone to mind-changes are to be eschewed and establish, in some contexts, a link between counter-inductive behaviour and mind changes---see, e.g.,  \citep{kelly2011simplicity} and \citep{Lin:2018aa}.} 

The results developed above presuppose that we are operating in an austere setting---one in which we countenance arbitrary (computable) data streams or data streams generated by sampling from arbitrary (computable) probability measures. In more tightly constrained settings, learning becomes tractable---e.g., if one knows that the data stream is  generated by a Bernoulli measure, then it is a straightforward task to use the data to successfully estimate the relevant parameter. But this observation illustrates rather than undercuts the perspective of the preceding paragraphs, making the point that although a universal learning algorithm is an impossibility, learning becomes possible when sufficiently strong presuppositions are in play. Of course, one would ultimately like to know more about where the boundaries lie of the class of learning problems in which failure is typical and of the class of learning problems in which success is typically achievable.\footnote{See \citep{fortnow1998relative} for some results of this kind for the problem of identification of sequences.}

The results developed above are absolute in the sense that they do not presuppose the choice of a privileged measure on Cantor space or on the space of probability measures on Cantor space. But most of them \emph{do} depend on the choice of topology. For the results concerning (weak) learning of sequences by extrapolators, this is not very worrying. In the vast majority of applications in statistics, economics, and computer science, the space of binary sequences is equipped with the product topology. And with good reason: this topology can be thought of as the topology of point-wise convergence and motivated by thinking of binary sequences as encoding real numbers in the usual way.  The situation is not quite as straightforward with the space of probability measures on Cantor space. Certainly, the weak topology is extremely natural---but it is only one of several natural options.  So it is natural to wonder whether the intractability of our learning problems would hold under other reasonable choices of topology.

\newpage

\bibliographystyle{elsarticle-num.bst}

\bibliography{lunch}

\end{document}